\documentclass[12pt, nofootinbib,preprint]{revtex4-1}
\usepackage{amsmath,amssymb,pxfonts}
\usepackage{verbatim,graphicx}
\usepackage{ytableau}
\usepackage{color}
\usepackage[all]{xy}

\newcommand{\ket}[1]{| #1\rangle}

\newcommand{\braket}[2]{\langle #1| #2\rangle}
\newcommand{\BR}{\mathbb{R}}
\newcommand{\BZ}{\mathbb{Z}}

\newcommand{\BC}{\mathbb {C}}

\newcommand{\ECR}{E$\chi$RS }
\newcommand{\MFf}{{\mathfrak f}}

\def\Tr{\textrm{Tr}}

\newcommand{\beq}{\begin{equation}}
\newcommand{\beqs}{\begin{equation*}}
\newcommand{\eeq}{\end{equation}}
\newcommand{\eeqs}{\end{equation*}}
\newcommand{\CO}{  {\cal O}  }

\allowdisplaybreaks

\begin{document}
\setlength{\unitlength}{1mm}
%%%%%%%%%%%%%%%%%%%%%%%%%%%%%%%%%%%%%%%%%%%%%%%%%%%%%%%%%%%%%%%%
\title{Extremal chiral ring states in AdS/CFT are described by free fermions for a generalized oscillator algebra}

\author{David Berenstein}
\affiliation {
Department of Applied Math and Theoretical Physics, Wilbeforce Road, Cambridge, CB3 0WA, UK\\and\\  Department of Physics, University of California at Santa Barbara, CA 93106}

\begin{abstract} 
This paper studies a special class of states for the dual conformal field theories associated with supersymmetric $AdS_5\times X$ compactifications, where $X$ is a Sasaki-Einstein manifold with additional $U(1)$ symmetries. Under appropriate circumstances,  it is found that elements of the chiral ring that maximize the additional $U(1)$ charge at fixed R-charge are in one to one correspondence with multitraces of a single composite field. This is  also equivalent to  Schur functions of the composite field. It is argued that in the formal zero coupling limit that these dual field theories have, the different Schur functions are orthogonal. Together with large $N$ counting arguments, one predicts that various extremal three point functions are identical to those of ${\cal N}=4 $ SYM, except for a single normalization factor, which can be argued to be related to the R-charge of the composite word. The leading and subleading terms in $1/N$ are consistent with a system of free fermions for a generalized oscillator algebra. One can further test this conjecture by constructing coherent states for the generalized oscillator algebra that can be interpreted as branes exploring a subset of the moduli space of the field theory and use these to compute the effective K\"ahler potential on this subset of the moduli space.
\end{abstract}

\preprint{DAMTP-2015-7}

\maketitle

%%%%%%%%%%%%%%%%%%%%%%%%%%%%%%%%%%%%%%%%%%%%%%%%%%%%%%%%%%%%%%%%
\section{Introduction}
\label{S:Introduction}

Half BPS states in ${\cal N}=4 $ SYM for the gauge group $U(N)$ are described by free fermion droplets in a two dimensional plane \cite{Berenstein:2004kk}. These fermions are in the lowest Landau level of a constant magnetic field that threads the plane. These configurations can be described in the thermodynamic limit in terms of an incompressible fluid in two dimensions. Surprisingly, the dual supergravity description of these states is the same: incompressible fluid droplets on a plane \cite{Lin:2004nb}, and one can argue that the collective coordinate quantization of these solutions reproduces the free fermion picture \cite{Mandal:2005wv,Grant:2005qc} . These states are constructed out of multitraces of a single complex scalar field, which we call $Z$. This is equivalent to writing the set of states in terms of Schur functions of $Z$ \cite{Corley:2001zk} , which are in one to one correspondence with Young diagrams. The fermionic description also follows from the description of the complete orthogonal set of states in terms of Young diagrams, where they can be put in one to one correspondence with Slater determinants. 

This construction of half BPS states in ${\cal N}=4 $ SYM can be generalized to other groups \cite{Caputa:2013vla} (see also \cite{Mukhi:2005cv}) and similarly it is found that the half BPS states are  described by free fermions in a suitable quotient of the plane. It is natural to try to extend this idea to orbifolds (and other orientifolds) of ${\cal N}=4 $ SYM. 
BPS states with less supersymmetry are not described by free fermions, although it can be argued that they are described by a holomorphic quantization of the moduli space of vacua (the quantization of the common eigenvalues to the matrices $X,Y,Z$), and that the volume of the gauge orbit can play a similar role to the fermion repulsion \cite{Berenstein:2005aa}. Such models, based on commuting matrix models, seem to get part of the physics correctly in much more general setups \cite{Berenstein:2007wi, Berenstein:2007kq}, but they have trouble calculating in a systematic manner. For example, now that the energies of some open string states between BPS D-branes  have been understood exactly to all orders in perturbation theory \cite{Berenstein:2013eya,Berenstein:2014isa} ( the results follow from understanding the central charge extension of the ${\cal N}=4 $ SYM spin chain in detail \cite{Beisert:2005tm,Dorey:2006dq}, but for the case of open strings), the approximations that led to the conjectural description of BPS states in  \cite{Berenstein:2005aa} do not seem to have any sensible way to reproduce them. Indeed, the free fermion picture has become even more important in describing the correct physics for these D-branes as coherent states \cite{Berenstein:2013md, Berenstein:2014zxa}.
Another problem that has arisen recently is that the geometry arising from commuting matrix models of many matrices seems to be renormalized and essentially collapses when curvature corrections to the effective dynamics on moduli space are included \cite{Filev:2014qxa}. This suggests that many of these ideas  on the dual geometry for gauge theories being based on an approximately commuting matrix model should be reformulated.

The success of the free fermion description of half BPS states and their excitations in the gravity dual gives hope  that maybe there are other situations in which a properly generalized free fermion description is possible. The purpose of this paper is to advance a set of conjectures on when this might be the case.

If one looks at the properties of the half BPS states in ${\cal N}=4$ SYM, they belong to the chiral ring and are highest weight states of the $SO(6)$ R-symmetry. When considering orbifolds with ${\cal N}=1$ SUSY that leave this sector invariant, one can check that the corresponding states are also in the chiral ring and they maximize an additional $U(1)$ charge which is not just the R-charge of the ${\cal N}=1 $ SCFT, and which is kept fixed. These can also be described in terms of free fermions on a quotient of the plane. This will be seen to be a consequence of more general arguments presented in this paper.

The main idea of this paper is to propose that there is a class of states in more general four dimensional CFT's (that are not just orbifolds) with a gravity dual that can be described in terms of a free fermion system in two dimensions on an appropriate cone. These are going to be states that are in the chiral ring and that for fixed $R$-charge maximize an internal $U(1)$ symmetry charge which is not the R-charge of the ${\cal N}=1$ theory. Because of this property we shall call any such state an extremal chiral ring state (E$\chi$RS) . There are additional conditions that this $U(1)$ symmetry satisfies that needs to be included as part of the definition of what is required to have \ECR. Mainly, these conditions ensure that the set of such extremal states can be described in terms of multi-traces of a single composite field $\tilde Z$, or alternatively, Schur functions of the corresponding composite field. Geometrically, the condition that is required is to have the $U(1)_R$ and the extra $U(1)$ vector field on the Sasaki-Einstein manifold be parallel on a single fixed circle.

The basic idea for the free fermion realization is that when writing the multitrace states in terms of  Schur functions  in theories with a free field limit, they are orthogonal. Indeed, 
they represent the free fermions directly. In this paper it is shown that assuming this orthogonality of  states labeled by Young diagrams  plus large $N$ counting for more general non-free theories is sufficiently restrictive to make a large class of predictions that can be tested in supergravity, and that at least to the first two orders in a $1/N$ expansion they coincide with a free fermion system. 

Examples for \ECR  arise in toric field theories, where there are a lot of additional $U(1)$ symmetries. These theories are non-free, as the chiral fields have non-trivial anomalous dimensions. The standard example is the Klebanov-Witten theory \cite{Klebanov:1998hh}, where one can check with the techniques of Leigh and Strassler \cite{Leigh:1995ep} that
there is a non-trivial fixed point with a quartic superpotential and non-trivial anomalous dimensions for all fields. 
 The theory is not free  even though one can take a limit where the gauge coupling constants effectively vanish.

In these setups one can also find a one parameter family of related field theories where the superpotential terms are varied by phases. Only one combination of these phases is physical, and for roots of unity the resulting geometry is an orbifold with discrete torsion of the original theory. This is easiest to see in the original example of ${\cal N}=4 $ SYM quotients  itself \cite{Douglas:1998xa, Douglas:1999hq, Berenstein:2000hy} (this has been discussed for general toric theories in \cite{Benvenuti:2005wi}, see also \cite{Dasgupta:2000hn,Butti:2007aq}). In such orbifold geometries of non-trivial CFT's there are fixed circles where the effective dynamics has non-trivial twisted sector states. The elements of the chiral ring associated to such twisted sector states are examples of \ECR states. The effective theory of such twisted sector states localizes in $AdS_5\times S^1$, rather than a full ten dimensional geometry and can be argued to be almost independent of the details of the ten dimensional geometry, except perhaps for the radius of $S^1$. A particular case of this $AdS_5\times S^1$ geometry also arises from orbifolds with ${\cal N}=2 $ Supersymmetry, and  it has been argued that these give insight into the $(0,2)$ six dimensional conformal field theories \cite{Aharony:2015zea}. At weak coupling this twisted  sector in ${\cal N}=2$ can be described by free fermions, and one can also understand exact properties of the corresponding spin chain that computes anomalous dimensions, even when it is not integrable \cite{Gadde:2010ku}. Seeing that these special cases on $AdS_5\times S^1$ for special radii of the $S^1$ lead to free fermions suggests that in the more general setup we study, where the $S^1$ can end up having a different size, this can also be the case.

The paper is organized as follows. In section \ref{sec:CR} the basic properties of the chiral ring in conformal field theories is discussed. It is argued that the multiplication 
structure of the chiral ring is essentially trivial except for the norm of the states. Section \ref{sec:YD} shows how one builds states labeled by Young diagrams for more general theories. A particular example that is dealt with in detail is an abelian  orbifold of ${\cal N}= 4 $ SYM, and it is also shown in this example why the different states labeled by Young diagrams are orthogonal in the free limit.
This is used to show how \ECR are described by Young diagrams (Schur functions) by construction. In section \ref{sec:conj} the basic conjecture of orthogonality of Young diagram states is proposed and some evidence in favor of the conjecture is  given. In some examples this orthogonality can be argued based on enhanced global symmetries of limits of field theories that are believed to exist. 
 Next, in section \ref{sec:cons}, together with large N counting arguments,  the proposed orthogonallity is used to bootstrap the leading $N$ and $1/N$ correction to the  norm of the Schur basis states. This is used to show that certain extremal three point functions of supergravity fields (to leading order in $1/N$)  coincide exactly with those of ${\cal N}=4 $ SYM except for a single constant. In section \ref{sec:FF} it is argued that this result coincides with a free fermion description for generalized oscillators to leading and subleading orders in $1/N$ and it is conjectured that this free fermion description is true to all orders. Various consistency checks are also performed. In particular the structure 
of D-branes as coherent states of the corresponding generalized oscillators is discussed, and from them the K\"ahler potential of a single D-brane is recovered. The paper ends with a discussion of the results and applications to the AdS/CFT program.

\section{Chiral ring states in 4D SCFT's}\label{sec:CR}

Let us consider a field theory in four dimensions with at least one supersymmetry.  A similar set of results can be argued in three dimensions with $N=2$ supersymmetry.
One can easily check \cite{Cachazo:2002ry} that if $\CO(x)$ is a chiral operator then we have that 
\begin{equation}
\partial_\mu \CO(x) = [P_\mu, \CO(x)]=\sigma_{\mu \dot \alpha \alpha} \{ \bar Q^{\dot \alpha},[Q^\alpha,\CO(x)]\}
\end{equation}
so that in the set of vacua of a field theory, the product of chiral fields is independent of the position of the operators
\begin{eqnarray}
\partial_{x_1^{{\alpha\dot \alpha}}}\langle \CO_1(x_1)\dots \CO_k(x_k)\rangle&=& \langle [\{ \bar Q^{\dot \alpha},Q^\alpha\},\CO_1(x_1)]\dots \CO_k(x_k)\rangle\\
&=&  \langle \{ \bar Q^{\dot \alpha},[Q^\alpha,\CO_1(x_1)]\}\dots \CO_k(x_k)\rangle\\
&=&0
\end{eqnarray}
where in the second line we have made use of $[Q^{\dot\alpha},\CO(x_i)]=0$ and the Jacobi identity. This is equal to zero on supersymmetric vacua (where $\bar Q$ acting on the left or right is zero) by integration by parts.

What this means is that we can specify a composite chiral operator $\CO_1(x_1)\dots \CO_k(x_k)$ without the position labels and this structure gives the set of chiral operators a ring structure. The ring structure is essentially trivial multiplication of operators by taking the position labels on the same position. When combined with conformal invariance, this statement is more powerful. 

Consider the superconformal algebra in four dimensions with $N\geq1$ SUSY. The commutation relations between the special superconformal generators and the $Q$ are schematically as follows (ignoring signs and factors of 2)
\begin{equation}
\{Q^I_\alpha, (S_J)_\beta\} \simeq \delta^I_J M_{\alpha\beta} +\delta^I_J\epsilon_{\alpha\beta} \Delta +\epsilon_{\alpha\beta}R^I_J\label{eq:SCA}
\end{equation}
where $M$ is a subset of the rotations (anti-self-dual in Euclidean signature), $\Delta$ is the dilatation operator, and $R^I_J$ is a generator of the R-charge. When we compactify the field theory on $S^3\times \BR$, we find that $S_{J}\simeq (Q^J)\dagger$ are adjoints of each other. Thus, we find that the left hand side of equation \eqref{eq:SCA} is a positive definite hermitian operator (it is a square). This produces an inequality for all states, where 
\begin{equation}
\Delta \geq R +S \label{eq:BPS}
\end{equation}
that is, the dimension of an operator needs to be greater than or equal to the R-charge plus the spin \footnote{We take a free chiral scalar field to have R-charge equal to one}.
When considering a chiral ring state, it is considered to be a chiral super primary. This means that it is annihilated by all of the $S$ and some of the $Q$ generators. 
For such states, the left hand side of \eqref{eq:SCA} will vanish, so the inequality \eqref{eq:BPS} is saturated. That is, in superconformal field theories we consider elements of the chiral ring to be such that their dimension is equal to their R-charge.

When we take the OPE of two such chiral ring operators $\CO_1,\CO_2$, we find that by $R$-charge conservation, the list of operators that can appear in the OPE on the right hand side have R-charge that is equal to $R_1+R_2$
\begin{equation}
\CO_1(0) \CO_2(x) \simeq \sum |x|^{\Delta - R_1-R_2}  \CO_{\Delta}(0)
\end{equation}
and positivity of $\Delta - R_1-R_2\geq 0$ ensures that the OPE is non-singular in the limit \footnote{Here this means that there is an absence of an inverse power law of $x$, so the limit exists. Not that the Taylor series is regular.}. Moreover, we have the trivial structure of the chiral ring that allows the definition of composite operators by trivial multiplication, that is
\begin{equation}
\lim_{x\to 0}\CO_1(0) \CO_2(x) = \CO_1(0)\CO_2(0)
\end{equation}
 In this sense, the OPE coefficients of the chiral ring seem to be completely trivial. The caveat is that we need to know the norm of the states in order to be able to determine properly normalized OPE coefficients. This is basically because to normalize the states we need information about the K\"ahler potential of the field theory. This is similar to the statement that Yukawa couplings can be computed in string theory, with the caveat that they are in a holomorphic normalization of the fields. Such Yukawa couplings would not be enough on their own to compute a scattering amplitude or a fermion mass. 
 
 What is interesting to consider is that the two point functions defining the Zamolodchikov metric are not parts of the chiral ring. Instead, one has that
 \begin{equation}
\langle \CO_i(0) \bar \CO_j(x) \rangle \simeq {\mathfrak H}_{ij}\delta_{\Delta_i, \Delta_j} \frac{1}{|x|^{2 \Delta_i}}
 \end{equation}
where the non-trivial information is encoded in the norm ${\mathfrak H}_{ij}$, which is a positive definite matrix. For each value of the $R$-charge, it is a finite dimensional matrix.

The objective of this paper is to make a case for a particular form of ${\mathfrak H}_{ij}$ for a subset of elements of the chiral ring. We have labeled these the extremal chiral ring states (\ECR).
The main conjecture will be that ${\mathfrak H}_{ij}$ is diagonal in a particular basis.

\section{Extremal chiral ring states are described by Young diagrams }\label{sec:YD}

Before we embark on a full description of extremal chiral ring states, it is best to start with an example where the rules for describing such states can be understood simply. We can then generalize the ideas to more general setups.
The idea is to look for generalizations of the half BPS states of ${\cal N}=4 $ SYM to orbifolds of such a theory. 

Consider an abelian orbifold field theory describing D branes on $\BC^3/\BZ_k$, where the $\BZ_k$ acts as 
$(\omega, \omega, \omega^{-2})$ on the $X,Y,Z$ variables and $k$ is odd. Such orbifold field theories are constructed by the method of images \cite{Douglas:1996sw} and make standard examples of the AdS/CFT correspondence \cite{Kachru:1998ys} (general $SU(3)$ quotients are discussed in \cite{Hanany:1998sd}). The field theory can be described by a quiver diagram with $k$ nodes (labeled $1, \dots k \mod(k)$), each with a gauge group $U(N_i)$ (where the values of $N_i$ are all equal to $N$). The $X,Y$ fields are in bifundamentals of the $(N_i, \bar N_{i+1})$, while the $Z$ fields are in the bifundamentals of 
$(N_i, \bar N_{i-2})$. This is depicted in the following graph.
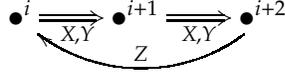
\begin{equation*}
\xymatrix{\bullet^i  \ar@{=>}[r]_{X,Y} &\bullet^{i+1}  \ar@{=>} [r]_{X,Y} & \ar@{->}@(dl,rd)[ll]_{Z}\bullet^{i+2}\\
&&&&}
\end{equation*}
It is convenient to introduce a formal associative algebraic structure so that $X=\oplus X_{i,i+1}$ etc, along the lines of \cite{Berenstein:2001jr}
(see also \cite{Berenstein:2009ay}). This algebraic structure is the path algebra of the quiver.
The algebra has additional orthogonal projectors that describe the nodes of the quiver diagram $\pi_i$ (here orthogonality means $\pi_i \pi_j = \pi_i \delta_{ij}$). Multiplication of fields in the algebra automatically includes index contractions that enforce matrix multiplication. This makes it convenient to build gauge invariant quantities without specifying index contractions in detail.
Under these conditions 
the fields $X_{i,i+1}= \pi_i X \pi_{i+1}$ can be recovered by acting with the projectors. Consider now the gauge invariant operator
$\Tr(Z^s)$. This can be written as
\begin{eqnarray}
\Tr( (\pi_1+\dots +\pi_k) Z^s)&=&\sum_i \Tr( Z_{i,i-2} Z_{i-2,i-4} \dots Z_{i-2s+2,i-2s})\\
&=& \sum_i \Tr(  Z_{i-2,i-4} \dots Z_{i-2s+2,i-2s}\pi_{i-2s} \pi_{i} Z_{i,i-2}) \label{eq:tr1}
\end{eqnarray}
and it vanishes unless $\pi_{i-2s} =\pi_{i} $, that is, unless $2 s$ is a multiple of $k$, or basically $s$ is a multiple of $k$ itself, because we chose that $k$ to be odd. 

This is a simple way 
to check that only some of the traces will survive. 
 In the dual gravity theory, these traces are  gravitons with angular momentum $s$   on the $S^5$ \cite{Witten:1998qj}, and only $s$ that are multiples of $k$ survive. That is, the spectrum of spherical harmonics on the $S^5$ is reduced to a subset that is invariant under the orbifold action \footnote{ For the case where $k$ is even above, there are two such traces that can survive. This is due to the fact that the corresponding supergravity quotient of the sphere $S^5/\BZ_k$ is not smooth. Therefore there are additional twisted sector states that can appear at the singularity (see for example the discussion in \cite{Gukov:1998kk,Berenstein:2000hy}).}. For this example the definition of a \ECR is that it is built only out of $Z$ so that it is a half BPS state in the covering theory. This property maximizes the charge that counts copies of $Z$ for fixed dimension of an operator (here all the $Z_{i,i-2}$ are required to have the same charge, which we set to $1$).

Our goal is now to show that in general such states built only out of $Z$ in  a gauge invariant way can be described uniquely by Young diagrams. The first obvious statement is that if one takes $\tilde Z_\ell=\pi_\ell Z^k \pi_\ell $, it is a loop that winds around the quiver using each $Z_{i,i-2}$ arrow only once, and that  it also begins at node $\ell$ and ends at node $\ell$. This is therefore a composite arrow that is in the adjoint of the $(N_\ell,\bar N_\ell)$ representation. Notice then that algebraically we have that $\Tr(\tilde Z_\ell^m)=\Tr(\tilde Z_q^m)$ for any $\ell, q$. This is due to the cyclic property of the trace. Thus, where we choose to begin in the quiver does not matter. Moreover, it is obvious from equation \eqref{eq:tr1} that
$\Tr(Z^{mk})=k \Tr(\tilde Z_\ell^{m})$ for all $\ell,m$.  Thus, expressing all gauge invariant polynomials of the fields  in terms of sums of products of   traces of powers of  $Z$ or in terms of sums of products of   traces of $\tilde Z_\ell$ is equivalent. 

There can be additional states that are not of this form. This is  because in general there can be dybaryon operators \cite{Gukov:1998kn}, which can not be written as products of traces. These use the $\epsilon$ invariant tensor of $SU(N)$. 
The correct conformal field theory in four dimensions has a gauge group which is a product of $SU(N)$ and dibaryon states need to be described to get a complete picture of the \ECR.  Describing states that have dibaryon charge is beyond the scope of the present paper. 
If we forego the use of the $\epsilon$ invariant tensor, the only available invariant tensor that can be used to make gauge invariants is $\delta^{i}_j$ and this naturally leads to matrix multiplication contractions.

Consider therefore a gauge invariant state built out of the collection of the $Z_{i,i-2}$ fields. If the state carries no dibaryon charge, then the number of the $Z_{i,i-2}$ is equal to the number of the $Z_{i-2,i-4}$ etc. The field $(Z_{i,i-2})^{j}_{\tilde \j}$ carries quantum  labels under two distinct gauge groups. A lower index $\tilde \j$ needs to be contracted to an upper index $\tilde \j$ in order to make a gauge invariant state (this is what we mean by using the $\delta^{i}_j$ invariant tensor). Only $(Z_{i-2,i-4})^{m}_{\tilde m}$ has such an upper index. Therefore the general gauge invariant state will start looking as combinations that are composite ``matrices'' $(Z_{i,i-2})^{j}_{\tilde \j} (Z_{i-2,i-4})^{\tilde \j}_{m}$. We keep on going by noticing that the index $m$ needs to be contracted as well. What we see is that in order to build gauge invariant states we end up invariably constructing objects that depend only on $\tilde Z_\ell$. Standard arguments \cite{Berenstein:2004kk} then show that we can only get sums of products of traces. And it follows that the allowed set of states are in one to one correspondence with Young diagrams \cite{Corley:2001zk}. This takes into account all the relations that appear in the set of traces when $N$ is finite.

As is well understood from the Schur-Weyl duality, a Young diagram for $U(N)$ determines a unique (unitary)  irreducible representation of the group $U(N)$. If $V$ is a vector space in the fundamental representation on which $U(N)$ acts by unitary transformation, a Young diagram with $s$ boxes determines a unique subspace of $V^{\otimes s}$ (with the induced norm from $V$) that exactly transforms as copies of such representation. We will call any such representation $R$.The representation $R$  is selected by doing weighted sums of  permutations of the factors $V$ in the tensor product. 
 Different diagrams determine different such representations and states in different representations are orthogonal.  
 
 The action of an element  $g\in U(N)$ on $V^{\otimes s}$ is given by $$g( v_1\otimes \dots \otimes v_s)= g v_1 \otimes g v_2 \otimes \dots \otimes g v_s$$ and this action is extended by linearity onto the whole $V^{\otimes s}$. This action commutes with the permutations acting on the labels $v_i$. The action can be extended to any element in $GL(N)$, although in this case $GL(N)$ does not preserve the norm. Furthermore, one can extend this action to the set of arbitrary $N\times N$ matrices by substituting $g$ by any such matrix. The operation of replacing $g$ by an arbitrary matrix commutes with the permutation of the vectors. Therefore, it also commutes with the (orthogonal) projections onto the different irreducible representations of the symmetric group, and therefore also the decomposition into the irreducible representations  $R$ themselves. Thus, on each such $R$, a matrix $M$ acting on $V$ defines a linear operation from $R$ to itself.
We can therefore compute the character of $M$ on $R$, $\chi_R(M)= \Tr_R(M)$. This is invariant under changes of basis in $V$ (which in turn implement a similarity transformation on $M$ by the action of $g$, $M\to g M g^{-1}$ ). The Schur basis of states, in analogy with 
\cite{Corley:2001zk} (see also \cite{Dey:2011ea, deMelloKoch:2012kv, Caputa:2012dg}), is defined by $\CO_R=\chi_R(\tilde Z_\ell)$ inside $V^{\otimes s}$, this includes a normalization factor from the multiplicity that $R$ might have inside $V^{\otimes s}$. 
 
We will now argue that these are orthogonal in the free field limit.  It turns out the proof is simpler in this case than in ${\cal N}=4 $ SYM, which depended heavily on the computation of the norm of the state \cite{Corley:2001zk}. The idea is as follows. 
Write the $\chi_R$ as a weighted sum over permutations, schematically of the form
\begin{equation}
\chi_R(\tilde Z_\ell)\simeq  \sum_\sigma \chi_R(\sigma)  (\tilde Z_\ell)^{\sigma [i]}_{[i]} \label{eq:chidef}
\end{equation}
and $\chi_R(\sigma)$ is a character of $\sigma$ in the representation $R$ of the symmetric group.

Let us unpack the $Z_{\ell, \ell-2}$ as they appear in the chiral ring operator $\CO_R$. We have that the upper indexes of $\tilde Z$ are carried exclusively by  $Z_{\ell, \ell-2}$. Thus, when we sum over permutations over the upper indices in $\sigma [i]$, we are summing
over permutations of the upper indices of $Z_{\ell, \ell-2}^{\otimes s}$. Since $Z_{\ell, \ell-2}$ is in the fundamental of $U(N)_\ell$, in the tensor product we are projecting onto the representation $R$ associated to $U(N)_\ell$ in the same way that it was done in the abstract space $V$ 
above. In the free field theory we can then think of $U(N)_\ell$ as a global  symmetry (acting Unitarily on the Fock space of the $Z_{\ell, \ell-2}$ fields) that is weakly gauged and requires a Gauss law constraint only when we include the other degrees of freedom beyond this. Under this symmetry, different unitary representations of $U(N)$ appearing in the Fock space of the fields $Z_{\ell, \ell-2}$ are orthogonal.
Hence, different Young diagrams give orthogonal states.
One can use this idea to work a smart version of orthogonality for the case of Young diagrams in the ${\cal N}=4 $ SYM. The idea is that in the zero coupling limit one can consider an enhanced $U(N)_L\times U(N)_R$ symmetry that acts on the raising operators as bifundamentals (this is a sub symmetry of the $U(N^2)$ dynamics that rotates the free  chiral fields into each other). The diagonal is then weakly gauged, but before we do that , the Young diagrams states already have to be orthogonal. This is because the upper indices have been projected into a particular irreducible of $U(N)_L$ determined by the Young diagrams, and different Young diagrams project onto different orthogonal  irreducibles (by Schur-Weyl duality).

Now, going back to the general case, consider  that because $Z$ are bosonic objects, we have the operator identity
\begin{equation}
(Z_{\ell, \ell-2})^{i}_{j}( Z_{\ell, \ell-2})^{i'}_{j'}=( Z_{\ell, \ell-2})^{i'}_{j'} (Z_{\ell, \ell-2})^{i}_{j}
\end{equation}
This means that in expressions involving index contractions, any permutation of the upper labels can be undone at the expense of permuting the lower labels instead. When we think carefully about what the weighted sum over permutations in equation \eqref{eq:chidef} is doing, we find that that the upper indices transform in representation $R_\ell$, and the lower indices in the conjugate representation of the next group $\bar R_{\ell-2}$  (if the objects were fermionic instead, one would have to transpose and conjugate the Young diagrams between upper and lower indices \cite{Berenstein:2004hw}). To form singlets, the same representation $R_{\ell-2}$ needs to be used for the upper indices of $Z_{\ell-2,\ell-4}$
(this is the only way to obtain a singlet in the tensor product of $\bar R_{\ell-2}\otimes R'_{\ell-2}$). Thus, choosing an origin for $\tilde Z_\ell$ plays no role in the final description of the Schur basis. Any such choice of origin in the quiver gives the same basis for the list of operators.

We have found an orthogonal basis of states. The norm of a state built this way will be computed by Wick contractions. Following the ideas of \cite{Dey:2011ea}, it is obvious what the answer for the norm of a state is going to be. Given a Young diagram fill in the boxes starting in the left upper corner with $N$, adding one when moving to the right, and subtracting one when moving down.  As an example consider a Young diagram with three rows of sizes $3,3,1$ as shown in \eqref{eq:y331}.
\begin{equation}
\ytableausetup{mathmode, boxsize=3em}
\begin{ytableau}
 N & N+1 & N+2 \\
 N-1 & N & N+1 \\
 N-2
\end{ytableau}\label{eq:y331}
\end{equation}
The norm of the corresponding state in the orbifold theory will be given by multiplying the labels on all the boxes and raising it to the $k$-th power.
\begin{equation}
\left|\ \ytableausetup{mathmode, boxsize=0.5em}
\ydiagram{3,3,1}\ \right|^2 = \left[N(N+1)(N+2)(N-1) N (N+1) (N-2)\right]^k
\label{eq:normy331}
\end{equation}
This is checked by realizing that the norm of the state has to be proportional to the dimension of the $SU(N)$ representation to the $k-th$ power. 

For simplicity we are abusing notation and describing the states themselves by the corresponding Young diagrams without any additional information. Complex linear combinations of the states are allowed, and multiplication in the chiral ring ends up being handled by the Littlewood-Richardson coefficients (exactly as in \cite{Corley:2001zk}), using the Young diagrams as the basis of states.
This procedure can be generalized easily to the case when the gauge groups have different rank \cite{Caputa:2012dg}, by multiplying the
results of these products of numbers when we decorate the Young diagrams with the ranks for each gauge group in the chain instead of a common one $N$.  This general result seems to agree with \cite{Pasukonis:2013ts}. We will not need these for what we will do in this paper. Dibaryons in general will modify the Young diagrams between the different gauge groups by adding extra columns of length $N$ that distinguish Young diagrams for the different letters building up $\tilde Z_\ell$. The reason we do not consider them in this paper is because we don't yet have a nice formula that describes the norm as in equation \eqref{eq:normy331}. One can conjecture that we obtain such an answer by multiplying the corresponding norms of the diagrams for each letter, but this needs to be checked.

From here, it is possible to compute the norm of trace states by using expressions of the kind
\begin{equation}
\Tr \tilde Z_\ell^3 = \ytableausetup{mathmode, boxsize=0.5em}
\ydiagram{3}-\ytableausetup{mathmode, boxsize=0.5em}\ydiagram{2,1}+\ytableausetup{mathmode, boxsize=0.5em}\ydiagram{1,1,1}
\end{equation}
as well as the overlaps between various trace combinations.
As is obvious from taking the leading powers of $N$, we find that a young diagram with $s$ boxes has norm $N^{sk}$ plus $1/N$ corrections. From here, the norm of $\Tr Z^s_\ell$ is $s N^{sk}$ and the corrections are of order $1/N^2$, just like one expects from planar diagram counting arguments.

Other abelian orbifolds with or without discrete torsion will work exactly the same way. The quivers for abelian orbifolds with discrete torsion are of  the same type as those without discrete torsion \cite{Berenstein:2000hy} (of the brane box type as in \cite{Hanany:1997tb,Hanany:1998it}), with all ranks being equal.

We are now ready for a definition of what it means for a field theory to be able to contain \ECR. First, for the purposes of this paper we restrict ourselves  to field theories with a well defined large $N$ limit (that can be described in principle by perturbative strings) that leads to an AdS dual of the form $AdS_5\times Y$, where $Y$ can be a stringy geometry.
 By choice we take this to be an oriented string theory, so that all fields in the dual field theory are oriented arrows ( of the type $(F,\bar F)$). For convenience all of our examples are drawn from orientable field theories, although it should be relatively straightforward to generalize to unoriented theories with a bit of care (similar to what is found in \cite{Caputa:2013vla}).

 The idea is that a conformal  field theory of this type should have an extra $U(1)$ charge, such that extremizing the $U(1)$ charge while fixing the $R$-charge of a chiral ring state results in a collection of states that can be uniquely described as multitraces of powers of a single composite arrow (a closed loop in the quiver) which will play the role of $\tilde Z_\ell$. The set of \ECR are classified by Young diagrams tautologically.  Again, we have assumed that the states in question contain no dibaryons. 
 This condition can be stated geometrically in the case that  $Y$ is a Sasaki-Einstein manifold with an additional $U(1)$ charge. If $Y$ is a quasi-regular Sasaki-Einstein manifold \cite{QR}, then it can be thought of as a circle fibration over a complex surface that might have orbifold singularities. 
 The restriction of charges on the states should mean geometrically that if we take a massless point particle moving in the geometry, in
  the optical limit all the states that are needed are supported at a  single point of the base of the fibration, 
that is, they are supported on a unique circle inside Y.  This is a condition of non-degeneracy. These usually are singular loci on the base of the fibration. 
This also ends up being the correct condition for non-regular Sasaki-Einstein manifolds, examples  can be found in \cite{Gauntlett:2004yd} (see also \cite{Martelli:2004wu} for more details). Up to this point, the singling out of \ECR says nothing about the dynamics, other than stating there is an additional $U(1)$ charge. Examples of such quiver field theories are well known \cite{Feng:2000mi,Benvenuti:2004dy,Franco:2005rj,Franco:2005sm}. The condition of no F-term relations suggests that the corresponding fields $\tilde Z_\ell$ wind around the dimer in a straight line where no moves that respect the R-charge of the composite field are allowed. Once the geometry is available, one can study long operators and relate them to classical trajectories on this circle \cite{Benvenuti:2005cz}

A prototypical example is the Klebanov-Witten conformal field theory \cite{Klebanov:1998hh}. The theory has two gauge groups $SU(N)_1\times SU(N)_2$, and four chiral fields $A_{1,2}$, $B_{1,2}$ in the $(\bar N, N)$ and $(N,\bar N)$ representations of the gauge group respectively. The role of $\tilde Z_\ell$ can be played by $Z_1=B_1 A_1$. This maximizes the charge that counts copies of $B_1$ and $A_1$. The corresponding states in the chiral ring are highest weight states of $SU(2)\times SU(2)$.
  Choosing only to maximize the number of $B_1$ is not enough, as one could have the combinations $B_1A_1$ and $B_1A_2$ appear. 
When considering the $F$-terms, one has that $A_1 B_1 A_2\simeq A_2 B_1 A_1$ as far as the chiral ring is concerned. This means that operators with composites $A_1 B_1 A_2$ and $A_2 B_1 A_1$ should generically mix and only one linear combination would 
survive. We would not know a priori which is the linear combination that survives.
When we take this into account, we find that the condition we need for the charge to be maximized is that there is no possible mixing due to F-terms relating a particular combination of fields to another one, because there is no degeneracy of composite words that can appear. Moreover, if we choose to only consider states that maximize the number of $B_1$ relative to the R-charge of the state for all values of the R-charge (not just at the level of traces),  we would end up only with dibaryon states, and no states that would appear as supergravity excitations in the dual theory.

\section{A conjecture for non-trivial theories with \ECR} \label{sec:conj}
 The main conjecture that we will make in this paper is that the  \ECR states associated to different Young diagrams are orthogonal states in the conformal field theory (that is, they are orthogonal in the Zamolodchikov metric). This is automatically true for field theories that are near a free field limit.
 This conjecture can be motivated in general as follows. The conformal  field theories that have a supergravity type IIB dual of the Freund-Rubin ansatz form, have a one parameter family of marginal deformations, which is characterized by changing the type IIB coupling constant in the dual supergravity theory. The main effect of this procedure is to change the radius of the geometric solutions in string units, but otherwise leave the supergravity solution essentially unmodified. 
 This in particular does not seem to affect any computation one would do in gravity in $AdS_5\times X$ as an effective low energy field theory.
 In particular, one can make the dynamical string splitting and joining in ten dimensions as weakly coupled as desired. In the field theory, the corresponding marginal deformation changes the field theory gauge coupling constants for the gauge groups. In the limit $g_s\to 0$, one has formally the property that  $g_{YM}\to 0$ (see \cite{Argyres:1999xu} for how this can be justified in terms of holomorphic invariants in some special sets of theories similar to the Klebanov-Witten theory).

  As such, even though the chiral fields can have a large anomalous dimension, one should treat the gauge interactions as if they are a weakly gauged global 
 symmetry. This would suggest that so long as one can imagine that the chiral ring fields appearing in $\tilde Z_\ell$ are quasi free \footnote{Quasi free means here that there are no F-term relations between the fields, and the chiral ring multiplication is like ordinary multiplication, but they do not have canonical dimension}, then the different Young diagrams states should be orthogonal. More precisely, so long as one can argue that this quasi-free limit produces an enhanced $U(N)_L\times U(N)_R$ symmetry for every $\tilde Z_\ell$, with $U(N)_{R\ell}\times U(N)_{L \ell'}$ weakly gauged to the diagonal, the enhanced symmetry would provide the desired orthogonality between states. One can also argue that when all gauge coupling constants go to zero simultaneously, the superpotential should go to zero at the same time. 
 Thus in the quasi-free limit where all the gauge dynamics can be considered a weakly gauged global symmetry, all of the limits with different super potentials should coincide. This is exactly true for free field theory limits, where the conditions of $\beta$ functions vanishing relate the gauge coupling constant to the Yukawa couplings. In this more general case  it probably needs to be assumed. Hence  orthogonality properties  survive to leading order in perturbation theory in the weakly gauged case for all the family of theories where one can argue that the corresponding states are in the chiral ring.

 Let us consider an example in which the global symmetry can be argued for. This is the case of the Klebanov-Witten theory \cite{Klebanov:1998hh}, one can start with an $N=2$ theory in the UV, with gauge group $SU(N)\times SU(N)$, and add mass terms to the scalar partners of the vector multiplets. There are two gauge coupling constants that can be varied independently, not only in the UV.  In the infrared  theory there is a exact marginal deformation corresponding to changing the relative values of the two coupling constants \cite{Klebanov:1998hh}. We can go to a limit where one of them is treated at zero coupling and the other one is at finite coupling.
The renormalization group  flow will take us to a non-trivial infrared fixed point for the second gauge theory, which is in the conformal window. That theory in the infrared would be an $SU(N)$ gauge field with  $2N$ flavors and a quartic superpotential. By familiar arguments, in the absence of a superpotential, it would have an $U(2N)\times U(2N)$ global symmetry \cite{Seiberg:1994pq}. The ${\cal N}=2$ coupling to the adjoint partner for the vector field reduces the symmetry to a diagonal $U(2N)$

The coupling to the super partner of the gauge field in the UV arises from a superpotential of the form
 \begin{equation}
m^2 \Tr(\phi^2)+  \sum \Tr( \phi Q^I_1 \tilde Q^I_1-\phi Q^I_2\tilde Q^I_2)\label{eq:massgen}
 \end{equation}
 and the mass term generates a quartic super potential in the infrared with a $U(2N)$ symmetry.

 The sign difference between $Q_{1,2}$  is due to the fact that some are fundamentals of $SU(N)$ and the others are antifundamentals when considered in terms of the UV $N=2$ theory. By a field redefinition, it can be put into more standard form where the $U(2N)$ symmetry is obvious. The field $\phi$ is neutral under this global symmetry, so when we integrate it out, the symmetry persists for the whole RG flow from the UV to the IR. This symmetry contains an $SU(N)\times SU(N)$ global symmetry subgroup, where a composite field like $\tilde Q^I_2 Q^J_1$ would transform as an $(\bar N ,N)$. 
 We can still add additional quartic couplings that respect this subset of the global symmetry. The idea is that we split the $2N$ of $U(2N)$ as a $N_1\oplus N_2$ under the subgroup that we want to preserve. Here we would have $Q_1$ charged as $N_1$, $Q_2$ as an $N_2$, $\tilde Q_1$ as a $\bar N_1$ and $\tilde Q_2$ as a $\bar N_2$.
 
 Indeed, subsets of the composite gauge invariant  mesons  of the form $Q \tilde Q$ transform as $M_{11}\simeq (N_1,\bar N_1)$ , $ M_{12} \simeq (N_1,\bar N_2)$ , $M_{21} \simeq (N_2,\bar N_1)$ and $ M_{22}\simeq (N_2,\bar N_2)$ from which we can make composites that are invariant under the $U(N_1)\times U(N_2)$ symmetry. These are terms of the form
 \begin{equation}
\alpha\Tr(  M^2_{11})+\beta\Tr(M_{12}M_{21})+\gamma\Tr(M^2_{22})
 \end{equation}
 all of which are expected to be marginal in the IR theory.   One can worry a bit about the construction of the theory, as the superpotential that is generated this way has additional terms that do not appear in the original Klebanov-Witten theory. However, general arguments exist to prove that the corresponding coupling constants are marginal \cite{Green:2010da}.
 Tuning $\alpha,\gamma$ we can get rid of the superpotential terms in \eqref{eq:massgen} that are not part of the Klebanov Witten theory, all the while preserving the global $U(N_1)\times U(N_2)$ symmetry (which will be gauged to a diagonal once we turn on the gauge coupling constant of the second gauge group). Indeed, the mass term above generates a particular linear combination of these terms.
 
 In this particular case one can  argue that there is a point in the conformal manifold where one of the gauge groups is at zero coupling and the other one is strongly coupled, and where there is moreover an enhanced $SU(N)\times SU(N)$ symmetry for which the argument presented in previous paragraphs provides exact orthogonality between states. Such a point still corresponds to weak string coupling, since the closed string coupling constant is given by
 \begin{equation}
 \frac 1 {g_s} \simeq \frac{1}{g_1^2}+\frac{1}{g_2^2} 
 \end{equation}
 The full superpotential usually acquires additional terms from integrating out the scalar partner of the weakly coupled theory. These  additional terms break the $U(N)\times U(N)$ symmetry to the diagonal. These can be varied independently of the gauge coupling constant and can be set  to be identically equal to zero in this limit, or proportional to the weakest coupling constant. It should be noted that in general this would correspond to a very stringy compactification and the supergravity approximation is not expected to be valid, even at finite $g_s$, but the theories should exist as superconformal fixed points in the same family.    At this point in moduli space, one has exact orthogonality of states classified by different Young diagrams. 
 
  A different approach would be to use the q-deformations of the superpotential \cite{Benvenuti:2005wi} combined with setting one of the gauge coupling constants to zero, which is also marginal.
 The superpotential would then be of the form
 \begin{equation}
 \Tr(A_1 B_1 A_2 B_2- q A_2 B_1 A_1 B_2)
 \end{equation}
At $q=0$ there is an enhanced global symmetry of the superpotential, where one gets an $SU(N)\times SU(N)$ at zero coupling for the gauge coupling constant at that node, which is then weakly gauged to the diagonal. Notice that all the \ECR survive the q-deformations, as the F-terms do not mix these states with others for any value of $q$.  This can also be seen in the study of giant gravitons in supergravity \cite{Imeroni:2006rb}.

This is exactly the theory constructed above by said marginal deformations by tuning the global couplings $\alpha, \gamma$ to zero in the total superpotential. A similar argument can be worked for other field theories where one can turn some gauge coupling constants off at the same time that one uses the $q$-deformation to check that there is an enhanced symmetry, this is done in conjunction with the additional $U(1)$ charge that determines that extremal states are present. 
The argument demands  a renormalization group trajectory from a weakly coupled UV to the IR is such that the enhanced global symmetry is present for the whole RG-flow.  This can be accommodated in practice in many examples.

 \section{Consequences of orthogonality}
 \label{sec:cons}

 Large $N$ counting arguments suggest a different basis of approximately orthogonal states, where the orthogonality of states is due to factorization (in the sense of matrix models)  in the infinite $N$ limit.
 The conjecture of orthogonality of states represented by Young tableaux when combined with large $N$ counting arguments will be shown to have many important consequences
 that can be tested directly in the gravity dual.  The conjecture makes unambiguous predictions for a wide range of supergravity observables, on theory by theory basis. Failure of passing a test would just mean failure of orthogonality of Young tableaux states for that one particular theory. They might still be approximately orthogonal however, a discussion that we will not pursue further in this paper.

To setup the large $N$ closed string Hilbert space of  \ECR,  different traces act as single particle raising operators of an approximate Fock space \cite{Witten:1998qj}. This  gives a different basis than the one produced by Young tableaux, but both give 
rise to the symmetric functions of the $N$ eigenvalues of the matrix $\tilde Z_\ell$, and there is a well known dictionary between them in terms of the Weyl character formulae (some of the simplest such relations are also known as Newton's identities or Newton-Giraud formulae). There are consistency requirements that need to be met in order for both basis to be orthogonal and approximately orthogonal simultaneously. We will study these consistency requirements now. 

The $1/N$ counting can be written in terms of generalized Feynman diagrams (the general counting is well explained  in \cite{Witten:1979kh}, see also \cite{Berenstein:2003ah} for a discussion of what it means to be an approximate Fock space). The rules are explained briefly in the appendix \ref{app:LNC}. Let us label 
\begin{equation}
t_s=\Tr(\tilde Z^s_\ell) 
\end{equation}
With this definition,  we have that 
\begin{equation}
|\prod t_i^{n_i}|^2= \prod n_i! |t_i|^{2 n_i} (1+O(1/N^2))
\end{equation} 
The leading term is the free Fock space result and is represented by free propagation of strings between the in state and the out state. Other overlaps between different states are zero to leading order, but can have $1/N$ corrections in general.
 The $O(1/N)$ corrections are organized diagrammatically from  basic vertex diagrams that respect addition of $R$-charge (these are like 1PI graphs with the $t_i$ as external legs).

A more interesting example is to consider
  \begin{equation}
\braket{t_s t_u}{ t_a t_b}\simeq |t_s||t_u||t_a||t_b|( \delta_{sa} \delta_{ub}+\delta_{sb} \delta_{ua}+N^{-2} \delta_{s+u,a+b} A_{s,u;a,b})
\end{equation} 
 which receives corrections both from disconnected diagrams (these are the contributions with $\delta$)  and from connected diagrams with four legs, represented by $A_{s,u;a,b}$. 
The $A$'s themselves are generically functions of the t'Hooft coupling $g_{YM}^2 N$ and they also have a $1/N$ expansion. The $A_{[\alpha],[\beta]}$, where $[\alpha],[\beta]$ are are multi-indices are symmetric
$A_{[\alpha],[\beta]}= A^*_{[\beta],[\alpha]}$, and they vanish unless $\sum_i \beta_i = \sum_j \alpha_j$ which encodes $R$-charge conservation (see appendix \ref{app:LNC}).
 
 Let us now use the orthogonality of Young diagrams states to determine the relations between the $t_i$, relative to $t_1$. We will do this for $i=2,3,4$.
Let us start with 
\begin{equation}
t_1 = \ytableausetup{mathmode, boxsize=0.5em}
\ydiagram{1}
\end{equation} 
and call $|t_1|^2= T$. 
Now, we get that 
\begin{equation}
t_1^2 =\ytableausetup{mathmode, boxsize=0.5em}
\ydiagram{2}+\ydiagram{1,1}
\end{equation} 
Using orthogonality of the two Young diagrams, plus our approximate Fock space description, we find that
\begin{equation}
|t_1^2|^2 =\ytableausetup{mathmode, boxsize=0.5em} |\ \ydiagram{2}\ |^2+ |\ \ydiagram{1,1}\ |^2\simeq 2T^2(1+ O(1/N^2))
\label{eq:t1sq}
\end{equation} 
Now, we also find that since
\begin{equation}
t_2= \ytableausetup{mathmode, boxsize=0.5em}
\ydiagram{2}-\ydiagram{1,1}
  \end{equation}
 then it follows that to leading order in $N$
 \begin{equation}
 |t_2|^2=\ytableausetup{mathmode, boxsize=0.5em} |\ \ydiagram{2}\ |^2+ |\ \ydiagram{1,1}\ |^2\simeq 2T^2\label{eq:t2}
 \end{equation} 
  That is, the norm of $t_2$ is completely determined to leading order in $N$ from knowledge of the norm of $|t_1|^2$, which we have called $T$.
  Furthermore, we find that
  \begin{equation}
  \braket{t_2}{t_1^2}= \ytableausetup{mathmode, boxsize=0.5em} |\ \ydiagram{2}\ |^2-|\ \ydiagram{1,1}\ |^2 \simeq N^{-1} \sqrt{2} |T|^2 A_{2;1,1} 
  \label{eq:t1sqt2ol}
  \end{equation}
 So that adding equations \eqref{eq:t1sq} and \eqref{eq:t1sqt2ol}, we find that
 \begin{eqnarray}
 \ytableausetup{mathmode, boxsize=0.5em}
 |\ \ydiagram{2}\ |^2= T^2  (1 +\eta/N+O(1/N^2))\\
 |\ \ydiagram{1,1}\ |^2= T^2 (1-\eta/N+ O(1/N^2) )
 \end{eqnarray}
 for some $\eta$.  Here, it is better to write $\sqrt 2 A_{2; 1, 1}= 2 \eta$, rather than the other way around. Our end goal is to compute the norm of all the Young diagrams to leading and subleading order in $1/N$.

 At the next stage, with elements of charge 3,  we have that 
 \begin{equation}
 t_1^3= (t_1^2) t_1 =  \ytableausetup{mathmode, boxsize=0.5em}(\ydiagram{2}+\ydiagram{1,1}\ )\ \ydiagram{1}=\ydiagram{3}+2\ \ydiagram{2,1}+\ydiagram{1,1,1}
 \end{equation}
 and the other two states are
 \begin{eqnarray}
 t_2 t_1&=&  \ytableausetup{mathmode, boxsize=0.5em}(\ydiagram{2}-\ydiagram{1,1}\ )\ \ydiagram{1}=\ydiagram{3}-\ydiagram{1,1,1}\\
 t_3&=& \ydiagram{3}-\ \ydiagram{2,1}+\ydiagram{1,1,1}
 \end{eqnarray}
The detailed study of these can be found in appendix \ref{app:3-4}, which also includes the study of diagrams with four boxes.
 
 An interesting fact that results from the computations in the appendix is that 
 \begin{equation}
 \braket {t_3}{t_2t_1} =  \braket{t_1^3}{ t_2 t_1}  \simeq N^{-1} |t_3||t_2||t_1| A_{3;2,1} = \frac{3  \times 2} N T^3 \eta \label{eq:A321}
 \end{equation}
 so that the leading order  term in $A_{3; 2,1}$ is not an independent quantity! Moreover, one finds that there is  a non-trivial consistency check with $1/N$ counting, namely that $ \braket {t_3}{t_1^3} \simeq O(1/N^2)$ as expected. 
  In general, one is starting to see an interesting pattern forming, with consistency conditions that are rather stringent.

The pattern appearing in the appendix \ref{app:3-4} becomes obvious when we compare with the prescription that leads to \eqref{eq:normy331}. If we decorate the same diagrams as in \eqref{eq:normy331} by omitting the factors of $N$ (these labels are the content of the boxes), we find 
 \begin{equation}
\ytableausetup{mathmode, boxsize=2em}
\begin{ytableau}
 0 & +1 & +2 \\
 -1 & 0 & +1 \\
 -2
\end{ytableau}\label{eq:y331dec}
\end{equation}
and then the norm of a diagram seems to be given to leading and subleading order by the formula
\begin{equation}
|Y|^2 = T^{ \hbox{\# boxes of } Y} \left(1 + \frac \eta N \sum_{\hbox{boxes}}(\hbox{label of box })^\prime  \right)
\end{equation}
where the prime here indicates the box labels with the $N$ stripped form the box.
This can be used to show that
\begin{equation}
|t_n|^2 = n T^n(1+O(1/N^2))\label{eq:t_n}
\end{equation}
and that 
\begin{equation}
\braket {t_n}{t_i t_j}=\delta_{n,i+j} \frac \eta N |t_n| |t_i| |t_j| \sqrt{( n) (i)( j)} \label{eq:3pt}
\end{equation}
of which equation \eqref{eq:A321} is a special example. It is obvious that imposing some of the relations that are due to planar counting at higher orders in $1/N$ will also produce  relations for some terms of order $1/N^s$ for all integer $s$. The precise study of these is beyond the scope of the present paper. Notice that so far we have not proved these relations.

The equation \eqref{eq:t_n} can be thought of as stating that the leading planar diagrams between traces are exactly as in a one matrix model. The factor of $n$ is the volume of the translational symmetry group on the string, which is related to the cyclic property of the trace. This cyclic property is usually related to the level matching constraints on the string dual \cite{Berenstein:2002jq}. Whereas equation \eqref{eq:3pt} states that the three point functions are completely determined in terms of one parameter, and up to scaling by this parameter they are functionally of the same form as the three point functions for a Gaussian complex matrix model.

A route to derive \eqref{eq:t_n} is as follows, by induction. First, we can assume that to leading order in $N$ all different multi- traces are orthogonal to each other, and every time we have Young diagrams with $m$ boxes, there is a unique new generator in the traces at this level. The new generator is $\Tr(\tilde Z^m)$. If any two different Young diagram states are compared, so long as each of them has a non-zero coefficient for $\Tr(\tilde Z^m)$, we have that 
\begin{eqnarray}
\ket{Y_1}&=& \vec \alpha_1 t^{[\alpha]} + a_1 \Tr(\tilde Z^m)\\
\ket{Y_2}&=& \vec \alpha_2 t^{[\alpha]} + a_2 \Tr(\tilde Z^m)
\end{eqnarray} 
where the $t^{[\alpha]}$ are the multitraces with $m$ copies of $\tilde Z$, for which we already know the norm. This is because at this order in $N$ we have a Fock space of states for each trace. The deformations away from the Fock space structure are suppressed by $1/N$ due to planar counting.  Here the $\vec \alpha$ are computed from equation \eqref{eq:chidef}, as for each permutation group element $\sigma$ we assign a unique multitrace operator in the contractions.

 From here we obtain that 
$a_1^* a_2 |t_m|^2+\braket {\alpha_1}{\alpha_2}=0$ where the term $\braket{\alpha_1}{\alpha_2}$ encodes the information we already know. From here, if the norm of traces  to order $m-1$ are known, the answer for the norm of $t_m$ is unique. This means that the norm can be read from any known solution of the equations, if one is available, For example, the single matrix model. 

For equation \eqref{eq:3pt} we can work similarly. Again, at each order $m$ there is one extra generator. But now to first subleading order in $N$ we have to deal with the 
coefficients $A_{m;n, m-n}$, which are different from zero for $m> n > 0$. There are $[m/2]$ in total. Again, the planar counting arguments tell us that the subleading order in $N$ from the multitrace part of the  norm has been already determined. We now have to count how many relations we get from orthogonality between pairs of different $Y$. This number is
${P(m)} \choose { 2}$ where $P(m)$ is the number of partitions, so long as ${{P(m)} \choose { 2}}\geq [m/2] $ and the set of equations for the triple functions are linearly independent, we .should get a unique answer for the $A_{m;n,m-m}$ determined only by the leading non-planar contribution $A_{2;1,1}$ or what we labeled $\eta$. Again, we can read this relation if were have one solution (like the free matrix model), but it has to be rescaled to an arbitrary $A_{2;1,1}$. 

Incidentally, for $\eta=1$ this reproduces exactly the set of extremal correlators in ${\cal N}=4 $ SYM \cite{Lee:1998bxa}. When we look at the orbifold examples of the previous section, $\eta$ is replaced by $k$ and the normalization of $T$ is $T= N^k$. Replacing $\eta=k$ for the subleading order  is straightforward from planar diagram computations. This is  because when we open up an object like $\Tr(\tilde Z^n$, we can do it in $k\ell $ places, but once we choose our origin, the origin of $\\Tr(\tilde Z^{m-n})$ needs to be compatible, and there are only $(m-n)$ such places. A similar statement is true for  $\Tr(\tilde Z^m)$, where there are only $m$ compatible places. From here, the vertex end up being proportional to $k m n (m-n)$, while the norm squared  of the states in the free Fock space ends up being proportional to $m n (m-n)$. Dividing by the correct normalization factor, we find what we need.

As is well known, extremal correlators give rise to expressions of the form $0/0$, as discussed in detail in \cite{D'Hoker:1999ea} (see also \cite{Skenderis:2006uy}). The results of this paper would suggest that such extremal correlation functions are universal for \ECR and depend only on one parameter, which we have chosen to use $\eta$. In the orbifolds of the previous section, this is measured by $\eta=k$, the R-charge of the word $\tilde Z_\ell$ (here we normalize a free field to have charge $1$). We conjecture this behavior for all such cases. That is, we conjecture that 
\begin{equation}
|Y|^2 = \prod_{\hbox{boxes}} (\hbox{labels of boxes})^{R_{\tilde Z_\ell}}  \label{eq:conjec}
\end{equation}

 In general, this R-charge can be determined by a-maximization \cite{Intriligator:2003jj}, so one has now a full conjecture about extremal correlators for a large class of Conformal Field Theories that can be tested. Such a conjecture gives a result that is invariant under toric duality, which in general is a form of Seiberg duality where the ranks of the gauge groups don't change \cite{Beasley:2001zp, Feng:2001bn} (See also \cite{Franco:2005rj}), and the R-charge of any composite word is the same as in the Seiberg dual theory. 

Notice that the conjecture stated in equation \eqref{eq:conjec} is very strong, as it details expansions to all orders in $1/N$. Although we will present evidence for the conjecture in the next section, we will take it more as an example of an all-order solution with the right leading $1/N$ expansion determined by one parameter $\eta$, rather than {\em the unique solution} itself. Such uniqueness would have to be proven and having other possibilities is not ruled out by the arguments we have made so far. We will see that  there are in principle other possibilities that solve the leading $1/N$ problem but are qualitatively similar to the above.

\section{Free fermions for a generalized oscillator}\label{sec:FF}

As described in the previous section, a prototype for a function that measures the norm of any Young diagrams is to take the result for the norm of a diagrams in ${\cal N}=4$ SYM and to raise it to the power $\eta$. 
This is a much stronger conjecture than just orthogonality of the Young diagrams states would require: it includes corrections to all orders in $1/N$, not just the leading order correction.
I will now present evidence for this stronger conjecture, but I will also weaken the form appearing in equation \eqref{eq:conjec} to show how other possibilities can arise.

The first claim that will be made about the  norms in equation \eqref{eq:conjec}, is that they describe a set of $N$ free fermions on a generalized oscillator algebra. We will take this more general statement as a conjecture for the form of the solution of the leading $1/N$ equations.
Our generalized oscillator will be described essentially by a generalized raising operator
\begin{equation}
a^\dagger \ket n = f_{n+1} \ket n
\end{equation}
where the $f$ are real positive numbers and the labels $n$ are integers starting at zero.
The generalized oscillator is described in detail the appendix \ref{app:GOA}. We will make the statements below for this more general case and we will see how to specialize the general oscillator for the example we are arguing for.
 The set of states of the general oscillator itself are labeled by integers (an occupation number), and a complete collection of free fermion states on such a Hilbert space are in one to one correspondence with 
choosing $N$ different number occupation states in $\cal H$ and writing a Slater determinant wave function. If we use the total number operator 
\begin{equation}
\hat N_{tot}= \sum_i \hat N_i
\end{equation}
as a Hamiltonian, the ground state is defined by having the minimal occupation number $\hat N_{tot, 0}= \sum_{i=0}^{N-1} i = N(N-1)/2$ 

Given a set of the $N_i$, we construct a Young diagram by first ordering the $N_i$ in decreasing order. This does not change the state (except for perhaps a sign). We associate to this object a Young diagram with rows of size $N_i - (N-i)$. This is the explicit map between free fermions for the half BPS sector of ${\cal N}=4 $SYM and Young diagrams \cite{Berenstein:2004kk}. The norm of the Young diagrams states will be given by 
\begin{equation}
\prod_{\hbox{boxes}} |f_{\hbox{label of box }}|^2\label{eq:matmod}.
\end{equation}
These are Slater determinants of the form
\begin{equation}
Y=N_0\frac 1{\sqrt N!} \det 
\begin{pmatrix}(a_1^\dagger)^{N_1} & (a_1^\dagger)^{N_2} &\dots& (a_1^\dagger)^{N_N})\\
(a_2^\dagger)^{N_1} & (a_2^\dagger)^{N_2} &\dots & (a_2^\dagger)^{N_N})\\
\vdots & \vdots &\ddots &\vdots\\
(a_N^\dagger)^{N_1} & (a_N^\dagger)^{N_2} &\dots & (a_N^\dagger)^{N_N})
\end{pmatrix} (\ket 0)^{\otimes N}
\end{equation}
where $N^{-1}_0=\prod_{i<N} G^{1/2}_k$, with the ground state $\ket 0$ defined by
\begin{equation}
\ket 0 \simeq N_0\frac 1{\sqrt N!} \det 
\begin{pmatrix}(a_1^\dagger)^{N-1} & (a_1^\dagger)^{N-2} &\dots& 1\\
(a_2^\dagger)^{N-1} & (a_2^\dagger)^{N-2} &\dots & 1 \\
\vdots & \vdots &\ddots &\vdots\\
(a_N^\dagger)^{N-1} & (a_N^\dagger)^{N-2} &\dots & 1
\end{pmatrix} (\ket 0)^{\otimes N}
\end{equation}
The set of $a^\dagger_i$ commute with each other. This is how one builds the tensor product of multi-particle states, and then the Slater determinant selects the completely antisymmetric wave functions that define Fermi statistics. The formula above should be familiar from the study of matrix models with orthogonal polynomials \cite{Itzykson:1979fi,Gross:1989aw}. Notice that since in our case we are using a complex representation for the wave functions and eventually their coherent states, then the $(a^\dagger)^n$ should be identified with the monomial $z^n$, and $a^\dagger$ with multiplication by $z$. Presumably a rotationally invariant measure exists that reproduces the values of $f$ above. That should define implicitly a solvable matrix model. 

The set of norms, like those  in equation \eqref{eq:normy331} are a special case of this construction with $f_s=s^{k/2}$. 
Also, one can show that this is consistent with taking 
\begin{equation}
\Tr (\tilde Z_\ell^s) \simeq \sum_{i\leq N} (a^\dagger_i)^s
\end{equation}
We can then interpret the raising operators for the fermions $a^\dagger_i$ as (operator-valued)  eigenvalues of $\tilde Z_\ell$. In this Hilbert space they are automatically fermions. This generalizes the ideas in \cite{Berenstein:2004kk} in a straightforward way, where the fermonic character is actually computed from first principles. Here, it is generally assumed, but our result for the norms already proves it for the case of orbifolds of ${\cal N}=4 $ SYM. It is expected that  this can be derived directly from the corresponding matrix model of bifundamentals by taking the effective action on eigenvalues carefully.

A simple example to understand the pattern is the following
\begin{eqnarray}\ytableausetup{mathmode, boxsize=0.5em}
\Tr(\tilde Z_\ell)^2\ket 0&= &((a_1^\dagger)^2+2 a_1^\dagger a_2 +(a_2^\dagger)^2+\dots)\ket 0\\
&=& \left( \ydiagram{2}+ 2\ \ydiagram{1,1} - \ydiagram{1,1} \right) \ket 0\\
&=&  \left( \ydiagram{2}+ \ \ydiagram{1,1}\right) \ket 0
\end{eqnarray}
The term $(a_1^\dagger)^2$ on the fist factor of the Slater determinant takes $(N_1)_0=N-1$ to $(N-1)+2$. The second one takes the same leading  terms to $(N_1)_0\to (N_1)_0+1$ and 
$(N_2)_0\to (N_2)_0+1$. The third term acts on the second column by taking $(N_2)_0\to (N_2)_0+2$. Here we see that this term is not in descending order, and can be converted to standard form by flipping the first two columns. This results in a minus sign from Fermi statistics, which is the third term on the second line. Indeed, this is how one generally derives identities like those that lead to equation \eqref{eq:t4} as an alternating sum of hooks. 

This shows that the norm we proposed and multiplication rule for diagrams is describing the norm on a system of free fermions for a generalized oscillator.
Existence of a large $N$ limit (independent of coupling constants, as we are effectively at zero coupling) requires in general that the oscillator algebra has a nice large $N$ limit for the ratio
\begin{equation}
\frac{ f_{N+1}}{f_N} = 1+\frac{\eta}N+\dots
\end{equation}
so that 
\begin{equation}
\frac{ f_{N+2}}{f_N}=\frac{ f_{N+1}}{f_N} \frac{ f_{N+2}}{f_{N+1}}= (1+\frac{\eta}N+\dots)  (1+\frac{\eta}{N+1}+\dots)\simeq   (1+\frac{2 \eta }N+\dots)  
\end{equation}
and so on.
Now, in general, taking the telescoping  product $\prod_{i=N}^{N+k}  f_{i+1}/ f_i$ we find that  
\begin{eqnarray}
\frac{ f_{N+k}}{f_N}&\simeq& \exp( \sum_{i=0}^k \frac \eta {N+i}) \\
&\simeq& \exp( \eta \log(N+k) -\eta \log(N) +\hbox{convergent})\\
&\simeq& (N+k)^\eta N^{-\eta} \exp\left[ b_1/(N+k)-b_1/N+O(1/(N+k)^2)\right]
\end{eqnarray}
So that $f_{N+k}$ asymptotes to a power of $N+k$ in the large $N$ limit. The simplest such function is a power of $N+k$ itself. Here we see that other solutions that describe free fermions and respect the $1/N$ counting are in principle possible \footnote{For this argument we have assumed that the generalized oscillator structure is independent of $N$. A more general behavior is possible, but our results so far are consistent with this simpler assumption.}. 

It is interesting to understand if this describes the set of all possible solutions to combining large $N$ counting and orthogonality of Young diagrams. Free fermion states  demand the orthogonality of Young diagrams, but the converse is not necessarily true, unless the large $N$ counting assumptions are very restrictive.
Notice that for the special case of  a power law, we find that the number operator for the generalized oscillator algebra of the appendix satisfies
\begin{equation}
\hat N= (a^\dagger a)^{1/\eta}
\end{equation}
The Hamiltonian of the field theory on $S^3$ is proportional to $\hat N$ plus a shift. If we choose the ground state to have energy $0$, we get exactly that the energy of a Young diagram state is proportional to the number of boxes. Free fermions typically appear when discussing matrix model quantum mechanics of one hermitian  matrix (see \cite{Klebanov:1991qa} for a review), so here we need to consider quantum mechanics of a holomorphic $\tilde Z$ and it's canonically conjugate momentum. Under the right circumstances they should commute (mostly this is due to the gauge constraint).  The canonical conjugate of $\tilde Z$ can be replaced by a function of the adjoint of $\tilde Z$ (namely $\tilde Z^\dagger$) and $\tilde Z$ itself . This suggests that one is dealing with a matrix model of normal matrices (see \cite{Chau:1997pr,Alexandrov:2003qk} for more information on such systems). The ground state of such matrix quantum mechanics would usually lead to a matrix model partition function that can  be described in terms of free fermions.
Moreover, one in general argues that these matrix models lead to distributions that are domains with  sharp boundaries in the thermodynamic limit \cite{Teodorescu:2004qm}. These distributions are obtained either by fixing the collective coordinate wave function of  the matrix quantum mechanics, or equivalently by changing the potential in the associated matrix model.

To understand why the power that appears in equation \eqref{eq:conjec} should be the R-charge, we need to consider D-brane states. There are two classes of such states that should be part of this class. Giant gravitons \cite{McGreevy:2000cw} and dual giant gravitons \cite{Grisaru:2000zn,Hashimoto:2000zp}. In general, states that are like giant gravitons will be described by holomorphic embeddings in the Calabi-Yau cone \cite{Mikhailov:2000ya}. It is the  second class of states (dual giants)  that we are interested in. These are described by geodesics on the Sasaki-Einstein space \cite{Martelli:2006vh}.
In the case of ${\cal  N}=4 $ SYM, the dual giant graviton states are described by Young diagrams that have only a single row \cite{Corley:2001zk}. The study of coherent states  along the lines of \cite{Caldarelli:2004ig,Berenstein:2014zxa} (see also the earlier work for studying giant gravitons \cite{Berenstein:2013md}) show that they can be understood in terms of the Coulomb branch of the theory where we take one eigenvalue and separate it from the origin to a finite distance (see also \cite{Hashimoto:2000zp} where this was argued based on
the fact that the dual giant describes a domain wall in supergravity and at the domain wall the effective value of $N$ changes by one). 

We need the same interpretation here. The point is that to get an expectation value for a field that can be  interpreted as a classical field, it must be understood as a  member of a set of coherent states.   The proper way of dealing with coherent states of a generalized oscillator is described in the appendix \ref{app:GOA}. That is, we want eigenstates of the lowering operator appearing in Slater determinants. It is easy to see that such states with one dual giant graviton will be given by
\begin{equation}
\ket \lambda\propto N_0\frac 1{\sqrt N!} \det 
\begin{pmatrix}{\mathfrak f}_1(\lambda )& (a_1^\dagger)^{N-2} &\dots& 1\\
{\mathfrak f}_2 (\lambda) & (a_2^\dagger)^{N-2} &\dots & 1\\
\vdots & \vdots &\ddots &\vdots\\
{\mathfrak f}_N(\lambda)  & (a_N^\dagger)^{N-2} &\dots & 1
\end{pmatrix} (\ket 0)^{\otimes N} \label{eq:sldetcs}
\end{equation}
where we have that $\ket \lambda = {\mathfrak f}(\lambda) \ket 0$ and where ${\mathfrak f}(\lambda)$ is a power series in the corresponding raising operators which describes the coherent state, defined as follows 
\begin{equation}
{\mathfrak f}_k(\lambda)=N_\lambda \sum_{n=0}^\infty \frac{\lambda^n}{G_n} (a_k^\dagger)^n
\end{equation}
this is the same as the series in the appendix \ref{app:GOA}, but using the basis $(a^\dagger)^n \ket 0$ rather than the orthonormal basis defined by the $\ket n$ states.

One can easily check that these states are eigenstates of the following elements of the operator  algebra 
\begin{equation}
\Tr(a^\ell) \ket \lambda = \lambda^\ell \ket \lambda\label{eq:eival}
\end{equation}
In general, one can write multi-coherent states for the various rows as follows
\begin{equation}
\ket{\lambda_1, \lambda_2, \dots}\propto N_0\frac 1{\sqrt N!} \det 
\begin{pmatrix} \MFf(\lambda_1)_1 &\MFf(\lambda_2)_1 &\dots& 1\\
\MFf(\lambda_1)_2 &\MFf(\lambda_2)_2 &\dots & 1\\
\vdots & \vdots &\ddots &\vdots\\
\MFf(\lambda_1)_N  &\MFf(\lambda_2)_N  &\dots & 1
\end{pmatrix} (\ket 0)^{\otimes N} 
\end{equation}
which are eigenstates of the single trace lowering operators
\begin{equation}
\Tr(a^\ell) \ket {\lambda_1, \lambda_2, \dots}= \sum _i \lambda_i^\ell \ket {\lambda_1, \lambda_2, \dots}
\end{equation}
These would be interpreted as vacuum expectation values in the Coulomb branch of the theory, similarly to how one does it in half BPS solutions in supergravity \cite{Skenderis:2007yb} (see also \cite{Skenderis:2006uy} for the general computation of 3-point functions). The main way to understand that such a state produces a classical configuration in the Moduli space of D-branes is to realize that the collection of numbers that appear in 
\eqref{eq:eival} associate a c-number to each element of the extremal states in the chiral ring, and these c-numbers give a representation of the chiral ring algebra (the c-numbers multiply for products of chiral ring states). As such, they define a point in the moduli space of vacua of the theory in flat space. Because non-extremal chiral ring states can be argued to carry other charges, we can consistently set them to zero as classical fields. The geometric representation of those points on the moduli space depends on the precise details of the superpotential of the theory, generically one imagines them as a Hilbert scheme of points on a noncommutative geometry \cite{Berenstein:2002ge}(see also \cite{Berenstein:2001jr}). 

These D-branes are going to be located in the special circle of the Sasaki-Einstein manifold that was described earlier in the paper. They  have a space like topology of $S^3$ expanding on $AdS$ and sit at  a point in the Calabi-Yau manifold moving along the R-charge direction. Since this is generically a singular point of the base of the Sasaki-Einstein manifold, the typical state will in general be in the moduli space of `fractional branes' (see \cite{Diaconescu:1997br}, see also \cite{Berenstein:2000hy}). If the Calabi-Yau manifold is smooth, then these are part of the D-branes in the bulk, but they are still confined to the special cone in moduli space of a single brane described by the special circle. Now let us analyze these states more carefully. As already discussed, they are in correspondence with points in the moduli space of vacua, that is, classical branes on the moduli space. As such, they are the closest analog to a classical solution of the field theory equations of motion in the full dynamics. Such classical field configurations are exactly what was part of the ansatz for dynamics that leads to emergent geometry in \cite{Berenstein:2005aa, Berenstein:2007wi, Berenstein:2007kq}. Here the corresponding structure is derived from the Young tableaux basis.

The state $\ket\lambda$ in the one oscillator Hilbert space is itself of the form
\begin{equation}
\ket \lambda = \sum \frac{\lambda^n}{(n!)^{\eta/2}}\ket n
\end{equation}
as in equation \eqref{eq:apl}. Notice that so long as $\eta>0$, the state is normalizable for all $\lambda$.
When put into the Slater determinants, it gets truncated to the large $n$ tail
\begin{equation}
\ket \lambda_T = \sum_{n\geq N-1} \frac{\lambda^n}{(n!)^{\eta/2}}\ket n
\end{equation}
For large enough $\lambda$, 
\begin{equation}
|\ket \lambda_T |^2\simeq |\ket \lambda |^2
\end{equation}
and the fluctuations on the effective number operator  $\hat N_{eff}=\hat N- (N-1)$ are small relative to $\hat N_{eff}$ itself. This means that the coherent states can be thought of as a classical solution in supergravity with a fixed energy. The classical energy is given by
\begin{equation}
R[\tilde Z_\ell] (|\lambda|^2)^{\eta^{-1}}- R[\tilde Z_\ell](N-1) \label{eq: gh}
\end{equation}
plus small quantum fluctuations. This is exactly the same behavior for dual giant graviton states that was derived for ${\cal N}=4 SYM$ in \cite{Berenstein:2014zxa}, starting from the open spin chain Hamiltonian computed first in \cite{de Mello Koch:2007uv} (this ansatz for dual giant graviton classical states was  guessed first in \cite{Caldarelli:2004ig}). The formula stated above can only be valid for values of $\lambda$ that produce positive energy. In other instances the ignored quantum fluctuations will dominate. This puts effectively a lower bound on $\lambda$ so that the D-brane lies outside the fermion droplet. Having a fermion over-density in the interior can be tied to the appearance of closed time like curves in the dual geometry \cite{Caldarelli:2004mz}.

Now, to the extent that the state $\ket\lambda$ is a classical state in the Coulomb branch with a large vev, the vacuum expectation values of \eqref{eq:eival} are very large. Even at weak coupling, we expect that being in the Coulomb branch can generate a mass gap for off-diagonal fluctuations that is much much larger than the size of the sphere (we are studying the field theory on $S^3\times \BR$ after all). In this limit, being in flat space or being on an $S^3$ does not matter.  The only scale that matters  in the system at such large vacuum expectation values is that vacuum expectation value itself. The energy per unit volume on the sphere is then determined by dimensional analysis.

This energy must arise from couplings to the curvature of the sphere times the appropriate quantity that makes sense in dimensional analysis. At the level of the effective action, we expect that for such a classical BPS solution
\begin{equation}
S_{eff} \simeq E T = \int d\Omega dt R F(\phi)
\end{equation}
where $F$ is a scaling function of dimension 2 (this is the correct dimensionality for a four dimensional field theory).
The reason for the proportionality to the Ricci scalar curvature is that in flat space the energy of such a state must vanish because it corresponds to a vacuum state of the conformal field theory with spontaneously broken conformal symmetry. Hence it must vanish when $R=0$.
Thus, the energy per unit volume must scale as
\begin{equation}
{\cal E} = E/Vol(S^3) = O(1) |\lambda|^{2/R[\tilde Z_\ell]}\label{eq: asym}
\end{equation} 
Comparing equations \eqref{eq: gh} and \eqref{eq: asym} tells us that the only possible way this will work is if $\eta=R[\tilde Z_\ell]$ (as we stated in the conjecture). At this point, this is another consistency test of the conjecture.

Notice that equation \eqref{eq: asym} can also be understood as computing the K\"ahler potential on the moduli space of a single brane. This is from understanding the effective classical action on moduli space for constant field configurations in equation  (3.16) of \cite{Berenstein:2007wi}, where terms of the form above are classically Weyl covariant only if $F$ is suitably chosen, and in particular proportional to the K\"ahler potential. The K\"ahler potential is therefore proportional to 
\begin{equation}
|\lambda |^{2/\eta}
\end{equation} 
The metric is therefore the metric of a cone with a deficit angle, and corresponds to a cone geometry over the special circle we singled out geometrically when describing the extremality condition geometrically.  This is a consistency condition in the sense that  the full moduli space of a single D-brane should be a cone geometry over a Sasaki-Einstein manifold base. The picture is now clear, the free fermion system corresponds to free fermions on a cone geometry.
This can also be derived from effective actions for collective coordinates as constructed in \cite{Berenstein:2013md}.

Another piece of evidence for the current proposal is that on taking a plane wave limit in supergravity along the special circle  (analogously to \cite{Blau:2002dy}, and whose field theory understanding was described in \cite{Berenstein:2002jq}) one generally finds that the plane wave resulting geometry is universal \cite{Itzhaki:2002kh,Gomis:2002km,Pando Zayas:2002rx}. Three point functions should therefore give in this limit the same result as in ${\cal N}=4 $ SYM, except for a finite volume normalization factor that depends 
on details of the size of the circle (this discussion can be found in \cite{Berenstein:2002sa}), and this is exactly what was derived in equation \eqref{eq:3pt}.

One can furthermore argue that the set of toric Conformal field theories under question admit a $\beta$-deformation, along the line of \cite{Berenstein:2000hy,Benvenuti:2005wi}. 
For rational roots of unity, these should be interpreted as orbifods with discrete torsion of the undeformed theory. This can be made very specific for the Klebanov -Witten theory \cite{Dasgupta:2000hn}. The different stringy geometries that can show up (either supergravity deformations or quotients) are related to each other by T-duality \cite{Berenstein:2000ux}.
The states we have discussed here survive for all values 
of $\beta$, and the D-branes should be though of generically as fractional branes on the quotient singularities. One can argue that at the same time that the gauge coupling constant goes to zero, one should have superpotential  coupling constants go to zero at the same time (the proof can be checked in perturbation theory \cite{Leigh:1995ep}, and generically one should expect the Yukawa couplings to be related to the gauge coupling constants).  To the extent that these field theories are the same in the zero coupling limit,  the value of $\beta$ should  be irrelevant for the quasi-free normalization of the states. One can then argue that the correlators should survive turning on the coupling constant infinitesimally, so the value of $\beta$ does not matter. For $\beta$ a root of unity associated to a high power $w^s=1$ for large $s$, the corresponding trace states that we need are all in the twisted sector (except for a $1/s$ small fraction). In this sense all the physics  of correlators we have singled out in supergravity localizes in the T-dual geometries to the special circle that was singled out geometrically, and the details of the transverse geometry should matter very little. This is an argument that the physics of interest is in some sense living on$AdS_5\times S^1$, rather than the full Calabi-Yau geometry. Considering the ideas presented in \cite{Aharony:2015zea}, this should be interpreted in terms of a non-trivial compactification of the $(0,2)$ theory on $AdS_5\times S^1$. What we have here gives a generalization to when the circle of the $S^1$ is different in radius than the AdS radius. 
  
\section{Discussion}\label{sec:D}

In this paper we have made the conjecture that Extremal Chiral ring states are not only classified by Young diagrams, but that such Young diagram states are exactly orthogonal to each other. The classification into Young diagrams is essentially a tautology because extremal chiral ring states are chiral ring states which are multitraces of a single composite field.
It is the orthogonality of states that has important consequences. It has also been argued that a natural stronger conjecture for such states is that Young diagram states are exactly described by free fermion wave functions of a generalized oscillator algebra. If we have a set of such free fermions, one can establish a dictionary between Slater determinants with the Young diagrams states that show the Young diagrams are orthogonal to each other. One should be careful on the meaning of this conjecture. The Hilbert space of free fermions on generalized oscillator algebras are all the same when considered just as  abstract Hilbert spaces of multi-particle states. They have the same number of states at each energy (level) and therefore they are formally equivalent. What distinguishes these spaces as quantum systems is the action of the oscillator algebra itself on the states. It is the norm of the states generated from the action of the generalized oscillator algebra on the ground state of the system that matters and distinguishes between them and defines in what sense one has a generalized oscillator algebra. 

 It is natural to speculate that orthogonality of Young diagrams states plus the validity of large N counting together might be enough to prove that the setup we have described here is equivalent to a collection of free fermions in this sense always. Our results so far are consistent with oscillators that are N-independent except for the free fermion hypothesis, but a more general outcome might be possible, as is done in double scaling limits of matrix models to obtain planar diagrams (see \cite{Brezin:1977sv}). This problem in mathematical physics is independent of its applicability to describe chiral ring states in conformal field theories. The problem of the chiral ring is a problem of physics. 
  
The important point of the conjecture in this paper  is that it predicts a number that certain extremal three point functions in supergravity are identical to those of the ${\cal N}=4 $ SYM theory up a single constant. These can be tested in supergravity.
 
 Moreover, we have found a simple set of solutions of the large N equations that might in principle be applicable to a large class of conformal field theories, characterized by equation \eqref{eq:conjec}. The conjecture for these solutions passes the simplest tests that could be devised. For example, two field theories that are related to each other by toric Seiberg duality give rise to the same set of norms for the Young diagram states, as they only depend on the R-charge of the states. Here the Seiberg duality \cite{Seiberg:1994pq} protects both the charge and the shape of the tableau, when we think of some of the nodes of the quiver as if they carry only global symmetry labels rather than gauge theory labels.
 
The conjecture itself can be thought of as an extension of ideas that are known to work in the case of the free field limit of ${\cal N}=4 $ SYM and its orbifolds. This is accomplished by noticing that the formula which depends on an integer $k$, equation \eqref{eq:normy331} still has a valid large $N$ limit when we analytically continue on $k$ for real $k$ rather than just for $k$ integer. The interesting question then is if this analytic continuation is useful in the study of the AdS/CFT correspondence. When $k$ is not an integer, one can not describe the system in terms of free field theories. Instead, one needs to resort to a non-trivial CFT, of which although there are plenty \cite{Benvenuti:2004dy,Franco:2005rj,Franco:2005sm}, there is only limited information that is known about them: the R-charges of the fields, the form of the super potential  and some of the superconformal deformations. Their conjectured AdS dual 

This conjecture also provides additional information on the non-renormalization of three point functions in ${\cal N}=4 $ SYM. To the extent that Young diagrams might be orthogonal even at finite coupling, the parameter $k$ (or $\eta$) that controls the physics should be such that the K\"ahler potential on the  moduli space of a brane in the bulk is non-renormalized (this is true for the moduli space of ${\cal N}=4 $ SYM theory). This forces $\eta=1$ and the three point functions are protected from weak to strong coupling, as expected from \cite{Lee:1998bxa}. This argument ties the non-renormalization of the 3-point extremal functions to the non-renormalization of the K\"alher potential on moduli space and seems to be different in spirit to the arguments on Harmonic superspace \cite{Howe:1998zi} (see also \cite{Heslop:2002hp,Dolan:2004mu} and references therein and see also \cite{David:2013oha} for the study of $\beta$-deformations). Notice that this is not expected to be true in three dimensional field theories of ABJM type \cite{Young:2014lka} (the corresponding field theories can be found in \cite{Aharony:2008ug}).

What should be obvious from our arguments is that the motivation for the validity of the orthogonality of Schur functions is dependent on an effective gauge coupling constant going to zero ( one can understand free field setups in very general cases \cite{Pasukonis:2013ts}). As such, one might argue that the SYM physics we need is strictly perturbative. Within this approximation, we have calculated the K\"ahler potential for a single brane, but one can easily argue based on our coherent state formalism, that this applies to multiple branes as well and the effective K\"ahler potential on the moduli space is a sum over terms that are block diagonal in the coordinates of the branes.  Indeed, the K\"ahler geometry is the  geometry of a symmetric product of a cone.
 This type of structure lands us squarely on arguments that have appeared in \cite{Berenstein:2007wi} ( particularly those that pertain to section 5). So long as one can argue that the effective field theory arguments are valid on the Coulomb branch of branes in the bulk (while still being realized with free fermions), one should be able to extrapolate these beyond zero coupling. However, one can also argue that in general one expects that when collections of fractional branes are involved, there are corrections to the metric and even perhaps the complex structure on moduli space  (this is usually exemplified by partial gaugino condensates \cite{Dijkgraaf:2002dh,Cachazo:2002ry}) . The K\"ahler potential is already renormalized in typical ${\cal N}=2 $ SYM theories and one might even destroy some of these fractional branches in some setups  \cite{Berenstein:2003fx}, but this last possibility seems to be outside the set of configurations that we have discussed. Indeed, it seems that the branches that are obtained are amenable to study by an effective Seiberg-Witten curve instead, very similar to what is found in \cite{Dorey:2002pq}.
 What might save us in general  is that we want to work at finite t'Hooft coupling where $g_{YM}^2 N$ finite, with perhaps only a few extra fractional branes that can be treated in the probe approximation.  These extra fractional branes are essentially locally free effective field theories  that are sufficiently similar to those with ${\cal N}=2 $ Supersymmetry away from the origin in moduli space. The Fermi exclusion principle prevents these few fractional branes from exploring the non-trivial singularity at the tip of the cone, so only the large vev region is explored. These branes will be generically weakly coupled with an effective coupling constant of order $g_{YM}^2$ which is tiny and then perturbative effective field theory can be used to make arguments.  This should all change when we take a sizable fraction of the branes away from the tip of the cone. Then, although pairwise effects of fractional branes are small, their collective effect might be large. Such large departures from the probe limit would be the ones that one would form by making a large LLM droplet on the cone, along the lines of  \cite{Lin:2004nb}. It is not clear that such supergravity solutions can be constructed, even for orbifolds.
 Already in the case of orbifolds  these effects associated to corrections to the form of the moduli space would be already apparent when one deals with collections of fractional branes that do not correspond to collections of bulk branes. 
 
A natural question is then to ask if the orthogonality of Young diagrams is only approximate (with order $1/N$ corrections), and the approximations get worse as the size of Young diagrams grows. This might need to be considered if the various supergravity tests of free fermions fail.

Assuming that the supergravity tests turn out positive, it is clear that one can then compute the norm of other states in situations with a lot of symmetry, like in the Klebanov-Witten field theory \cite{Klebanov:1998hh}. 
Given that information one would have a large collection of normalized correlation functions and OPE coefficients where one could hope to be able to perform conformal perturbation theory calculations. Such calculations   would describe stringy corrections to the sigma model of strings in the dual theory. In particular, one can imagine that  for the type of
arguments presented in \cite{Benvenuti:2005cz}, it would be possible to go beyond a schematic presentation to actually compute the detailed spin chain for near-chiral ring states form which a sigma model for strings could be derived.

\acknowledgements
The author would like to thank S. Ramgoolam for various discussions and especially N. Dorey for various discussions and Alexandra Miller for comments on an early draft of the paper. 
Work  supported in part by the department of Energy under grant {DE-SC} 0011702. The research leading to these results has received funding from the European Research Council under the European Community's Seventh Framework Programme (FP7/2007-2013) / ERC grant agreement no. [247252].

\appendix

\section{Large N counting}\label{app:LNC}

In this appendix we detail the large N counting rules that are required to make various estimates for correlation functions. This follows the counting rules  found in  \cite{Witten:1979kh} (following \cite{'tHooft:1973jz}) and supplemented by including symmetry considerations of conservation of the R-charge.

The first statement is that traces are approximately orthogonal \cite{Witten:1998qj}, and we have collections of traces of a single composite matrix $t_s=\Tr(\tilde Z^s_\ell)$.  We take these to be kets. We associate an external (initial state) leg to a graph with the label $s$ for each such trace. For computing norms, we associate bra states to the duals $\bar t_k=\Tr(\bar{\tilde Z}^k_\ell)$. A free propagator is associated with the norm $\braket{ \bar t_k}{ t_i}= \delta_{k,i} |t_i|^2$. The factors of $|t|_i$ are attached to  the external legs. The leading order in $N$ correlation functions are made of these (disconnected) free propagators. For example
\begin{equation}
\braket{t_s, t_t}{t_{\tilde s}, t_{\tilde t}}=\xymatrix{ s \ar@{-}[d] & t\ar@{-}[d] \\
\tilde s & \tilde t}+\xymatrix{ s \ar@{-}[dr] & t\ar@{-}[dl] \\
\tilde s & \tilde t} =\left( \delta_{s, \tilde s} \delta_{t,\tilde t}+\delta_{s, \tilde t} \delta_{t,\tilde s}\right) |t_s|| t_{\tilde s}| |t_t| |t_{\tilde t}|
\end{equation}
To each external leg with label $i$, we associate a degree $i$. For a multi-index $[u]$, we associate a size $|u|$ which is the number of different labels in $[u]$, and a degree which is the sum of the degrees of the labels of $[u]$.

Schematically these additional connected diagrams  are of the form
\begin{equation}
\braket{t_{[u]}}{t_{[v]}}=\xymatrix{ [u] \ar@{=}[d]& \\
\bullet &A_{[u];[v]}\delta_{\deg[u], \deg[v]} 
\\ [v]\ar@{=}[u] &} =\left(\frac{ A_{[u];[v]}}{N^{|u|+|v|-2}}\right)\delta_{\deg[u], \deg[v]} |t_{[u]}|| t_{[v]}|
\end{equation}
where the labels ${[u]}, {[v]}$ denote multi-indices and the double lines indicate contractions of the external vertices into the vertex $A_{[u];[v]}$. Objects like $ |t_{[u]}|$ multiply the norms over all external legs in ${[u]}$. 

 The final answer for a correlator is a sum over all connected and disconnected contractions of the top labels and bottom labels with the corresponding factors of $N$.
The simplest such non-trivial correlation is the following:
\begin{equation}
\xymatrix{ &t_s \ar@{-}[d]& \\
& \bullet & A_{s; a,b} \delta_{s, a+b}\\
t_a \ar@{-}[ur] && \ar@{-} [ul] t_b} \ \ \ \ = N^{-1} A_{s; a,b}\delta_{s, a+b}|t_a||t_b||t_s|
\end{equation}

Similarly, one finds that
  \begin{equation}
\braket{t_s}{ t_a t_b t_c}= N^{-2} |t_s| |t_a| |t_b||t_c| \delta_{s, a+b+c} A_{s;a,b,c}
\end{equation} 

Each of the quantities $A_{[u],[v]}$ are formal power series in $1/N^2$. 

\section{Norms of Young diagrams for 3, 4 boxes}\label{app:3-4}

The convention in this paper is that $t_i = \Tr(\tilde Z^i)$. Moreover, the approximate Fock space structure detailed in appendix \ref{app:LNC} implies that 
$|\prod t_i^{n_i}|^2= \prod (n_i !) |t_i|^{n_i}(1+O(1/N^2)$. We have that $|t_1|^2=T$. From the results of equation \eqref{eq:t2}, we find that $|t_2|^2= 2 T^2+O(1/N^2)$, and moreover
$\braket{t_2}{t_1^2}= 2 T^2 \eta/N+O(1/N^3) $. To proceed further, we need to study diagrams with three and four boxes respectively, and this includes traces up to $t_3, t_4$ respectively.

For three boxes, we have the states
 \begin{eqnarray}
 t_1^3&=& \ydiagram{3}+2\ \ydiagram{2,1}+\ydiagram{1,1,1}\\
 t_2 t_1&=&  \ytableausetup{mathmode, boxsize=0.5em}(\ydiagram{2}-\ydiagram{1,1}\ )\ \ydiagram{1}=\ydiagram{3}-\ydiagram{1,1,1}\\
 t_3&=& \ydiagram{3}-\ \ydiagram{2,1}+\ydiagram{1,1,1}
 \end{eqnarray}

Using $|t_1|^3 \simeq 6 T^3$ and $|t_2 t_1| ^2 \simeq 2 T^3$ plus order $1/N^2$ corrections and orthogonality of the Young diagrams we find that 
\begin{equation}
|\ \ydiagram{2,1}\ |^2 =  T^3(1+O(1/N^2))
\end{equation}
 If we now use the $A_{2;1,1}$ vertex and the corresponding Feynman graph counting, we get that
 \begin{equation}
 \braket{t_1^3}{ t_2 t_1} = |\ \ydiagram{3}\ |^2-\left |\ \ydiagram{1,1,1}\ \right|^2 = N^{-1} 3 |t_1|^4|t_2| A_{2;1,1}+ O(1/N^3)= \frac 3 N T^3 (2 \eta)+O(1/N^3)
 \end{equation} 
 from which it follows that 
 \begin{eqnarray}
 |\ \ydiagram{3}\ |^2 &=& T^3 (1+ 3\eta/N+O(1/N^2))\\
  |\ \ydiagram{1,1,1}\ |^2 &=& T^3 (1-3\eta/N+O(1/N^2))
  \end{eqnarray} 
 One then directly computes that
 \begin{equation}
 | t_3|^2 =  |\ \ydiagram{3}\ |^2 + |\ \ydiagram{2,1}\ |^2 +   |\ \ydiagram{1,1,1}\ |^2 = 3(T^3)(1+O(1/N^2))
 \end{equation} 
 so that again, one finds that to leading order $t_3$ is completely determined by the norm of $t_1$, and that corrections begin at order $1/N^2$. 
   
 At the next stage, we have the five states
 \begin{eqnarray}
 t_1^4 &=&( \ \ydiagram{3}+2\ \ydiagram{2,1}+\ydiagram{1,1,1}\  )\ \ydiagram{1}\\
 &=&  \ydiagram{4}+ 3\ \ydiagram{3,1}+2\ \ydiagram{2,2} +3\ \ydiagram{2,1,1}+  \ydiagram{1,1,1,1}\\
 t_2 t_1^2 &=& \ydiagram{4}+\ydiagram{3,1}-\ydiagram{2,1,1}-\ydiagram{1,1,1,1}\\
 t_3 t_1 &=& \ydiagram{4}-\ydiagram{2,2}+\ydiagram{1,1,1,1} \\
 t_2^2 &=& \ydiagram{4} -\ydiagram{3,1} +2\ \ydiagram{2,2}  -\ydiagram{2,1,1}+\ydiagram{1,1,1,1}\\
 t_4&=& \ydiagram{4} -\ydiagram{3,1} +\ydiagram{2,1,1}-\ydiagram{1,1,1,1}\label{eq:t4}
 \end{eqnarray} 
The approximate Fock space structure gives $5 \choose 2$ different possible overlaps (linear relations)  to determine the norm of the five Young diagrams, and the only other unknown at this step is the norm of $t_4$. Excluding $t_4$ gives ${4\choose 2}=6$ different overlaps, so that the system is over constrained at this step already and the norm of $t_4$ can be computed without using it in any other relation.
It is interesting to choose overlaps with few terms and simple linear combinations with few terms. For example, let $U= t_1^4-t_2^2$. It is easy to check that to leading and subleading order in $N$
\begin{equation}
|U|^2 = 16 \left|  \ydiagram{3,1} \right|^2 +16  \left|  \ydiagram{2,1,1} \right |^2= |t_1|^4 + |t_2|^2 +O(1/N^2) \simeq 4 ! T^4 + 2 ( 2T^2)^2 =32 T^4 \label{eq:rel1}
\end{equation}
because the overlap $\braket{t_2^2}{ t_1^4}$ is of order $1/N^2$.  The overlap between $t_2^2$ and $t_3t_1$ is also of order $1/N^2$, so we find that
\begin{equation}
\braket{t_2^2}{t_3t_1} = |\ydiagram {4} |^2+  \left| \ \ydiagram{1,1,1,1}\ \right |^2-2\left| \ \ydiagram{2,2}\ \right |^2 \simeq O(T^4/N^2) \label{eq:rel2}
\end{equation}
 Using the two relations \eqref{eq:rel1}  and \eqref{eq:rel2}  in 
 \begin{equation}
 | t_1^4|^2 =| \ydiagram{4}|^2+ 9|\ \ydiagram{3,1}\ |^2+4|\ \ydiagram{2,2}|^2 +9\left|\ \ydiagram{2,1,1}\ \right |^2+ \left|\ \ydiagram{1,1,1,1}\right|^2\simeq 24 T^4+O(1/N^2)
 \end{equation}
  we find that 
  \begin{equation}
  |\ \ydiagram{2,2}|^2 = T^4 +O(1/N^2)
  \end{equation}
 These relations alone are sufficient to show that 
 \begin{equation}
 |t_4|^2 = 4 T^4
 \end{equation} 
 Now, as a consistency check, we have that 
 \begin{equation}
 |t_3t_1|^2 =|\ \ydiagram{4}\ |^2 + |\ \ydiagram{2,2}\ |^2+\left|\ \ydiagram{1,1,1,1}\ \right |^2 = 3 T^4+O(1/N^2) =  |t_3|^2|t_1|^2(1+O(1/N^2))
 \end{equation} 
Consider that 
 \begin{equation}
 \braket{t_3t_1}{t_2t_1^2} = \frac 2N |t_1|^3 |t_3||t_2| A_{3;2,1} +O(1/N^3)=|\ \ydiagram{4}\ |^2 -  \left|\ \ydiagram{1,1,1,1}\ \right |^2  \simeq   2 \frac{3  \times 2} N T^4 \eta
 \end{equation}
 Which gets us to 
 \begin{eqnarray}
 |\ \ydiagram{4}\ |^2 &=& T^4\left[ 1+ \frac \eta N( 0+1+2+3) +O(1/N^2)\right] \\
\left  |\ \ydiagram{1,1,1,1}\ \right |^2 &=& T^4\left[ 1+ \frac a N( 0-1-2-3) +O(1/N^2) \right]
 \end{eqnarray}
 Similarly, the overlap
 \begin{equation}
 \braket{t_2^2}{t_2t_1^2} = \frac{2}N |t_2|^3|t_1|^2 A_{2;1,1}+O(1/N^3)
 \end{equation}
 together with previous results
 can be used to show that 
 \begin{eqnarray}
  |\ \ydiagram{3,1}\ |^2 &=& T^4\left[ 1+ \frac \eta N( 0+1+2-1) +O(1/N^2)\right] \\
\left  |\ \ydiagram{2,1,1}\ \right |^2 &=& T^4\left[ 1+ \frac \eta N( 0+1-2-1) +O(1/N^2) \right]
 \end{eqnarray}

\section{Generalized oscillator algebras and their coherent states}\label{app:GOA}

Consider an algebra defined by three operators on a Hilbert space $\hat N, a^\dagger, a$, such that $\hat N$ is self-adjoint with bounded from below (discrete) spectrum, $a^\dagger $ is the adjoint of $a$ and the list of the following commutator relations holds
\begin{eqnarray}
\ [\hat N, a^\dagger]&= & a^\dagger\\
\ [\hat N, a]&= & - a
\end{eqnarray}
It is easy to show that $\hat N$ commutes with $a^\dagger a$, that both are self-adjoint and therefore that they can be diagonalized simultaneously. We assume that $\hat N$ or $a^\dagger a$ can be expressed as functions of one another and that there is a unique irreducible representation of the algebra. This generalizes the harmonic oscillator algebra where 
$\hat N= a^\dagger a$ exactly. The setup here is more general than that in \cite{Arik:1973vg}.

 Let $\ket\alpha$ be an eigenstate of $\hat N$ with eigenvalue $\alpha$.
Then it is easy to check that 
\begin{equation}
\hat N  a \ket\alpha = a \hat N \ket \alpha - a \ket \alpha = (\alpha -1 ) a \ket \alpha\propto \ket{\alpha -1}
\end{equation}
That is, acting with $a$ lowers the eigenvalue of $\hat N$. Eventually we reach a lowest eigenvalue of $\hat N$, below which $ \ket{\alpha -k}$ would have an eigenvalue for
$\hat N$ which is below the bound. This state must identically vanish. This means, that the representation theory of the algebra has a (unique by assumption) lowest weight state  irreducible representation.

We can always choose the lowest weight state to have eigenvalue zero (noticing that shifting $\hat N$ by a constant does not affect the commutation relations).
Once we have this lowest weight state $\ket 0$, we can construct the representation by acting with copies of $a^\dagger$. Obviously $\ket 0$ is an eigenstate of $a^\dagger a$ with eigenvalue zero, and acting with $a^\dagger$ various times increases the eigenvalue of $\hat N$ by integer units.

Consider an orthonormal set of states $\ket n$.  Then it follows that 
\begin{equation}
a^\dagger \ket n = f_{n+1} \ket{n+1}
\end{equation}
where the $f_{n+1}$ can be chosen to be real and positive by rephasing $\ket{n+1}$. In a similar way, we find that
\begin{equation}
a \ket n = f_n \ket{n-1}
\end{equation}
It trivially follows that 
\begin{equation}
G_k=|(a^\dagger)^k\ket 0| = \prod_{i=1}^k |f_i|^2
\end{equation}
and we will call this quantity $G_k$. $G_0=1$ by convention. We will now define coherent states for the algebra as follows. A coherent state $\ket\lambda$ is an eigenstate of $a$. That is, we have that
\begin{equation}
a \ket \lambda = \lambda \ket \lambda
\end{equation}
The general form of such a state is of the following form
\begin{equation}
\ket \lambda = \sum_i a_i(\lambda) \ket i 
\end{equation}
and it can be shown easily that 
\begin{equation}
\ket \lambda =N_\lambda \sum \frac{\lambda^k}{\sqrt{G_k}} \ket k \label{eq:apl}
\end{equation}
where $N_\lambda$ is a normalization factor.
The state is normalizable if and only if
\begin{equation}
\sum_k \frac{|\lambda|^{2k}}{G_k}<\infty
\end{equation}
and that in turn requires that $G_k\neq 0$ for all $k$, so the representation is infinite-dimensional.
If we assume that  the series converges for some value of $\lambda$, it will converge also for all $|\lambda'|\leq |\lambda|$, so the domain of convergence is a disk in the complex plane.


\begin{thebibliography}{99}


%\cite{Berenstein:2004kk}
\bibitem{Berenstein:2004kk} 
  D.~Berenstein,
  ``A Toy model for the AdS / CFT correspondence,''
  JHEP {\bf 0407}, 018 (2004)
  [hep-th/0403110].
  %%CITATION = HEP-TH/0403110;%%
  %253 citations counted in INSPIRE as of 06 Jan 2015
  
  %\cite{Lin:2004nb}
\bibitem{Lin:2004nb} 
  H.~Lin, O.~Lunin and J.~M.~Maldacena,
  ``Bubbling AdS space and 1/2 BPS geometries,''
  JHEP {\bf 0410}, 025 (2004)
  [hep-th/0409174].
  %%CITATION = HEP-TH/0409174;%%
  %518 citations counted in INSPIRE as of 06 Jan 2015
  
 %\cite{Mandal:2005wv}
\bibitem{Mandal:2005wv} 
  G.~Mandal,
  ``Fermions from half-BPS supergravity,''
  JHEP {\bf 0508}, 052 (2005)
  [hep-th/0502104].
  %%CITATION = HEP-TH/0502104;%%
  %67 citations counted in INSPIRE as of 20 Apr 2015 
  
  %\cite{Grant:2005qc}
\bibitem{Grant:2005qc} 
  L.~Grant, L.~Maoz, J.~Marsano, K.~Papadodimas and V.~S.~Rychkov,
  ``Minisuperspace quantization of 'Bubbling AdS' and free fermion droplets,''
  JHEP {\bf 0508}, 025 (2005)
  [hep-th/0505079].
  %%CITATION = HEP-TH/0505079;%%
  %75 citations counted in INSPIRE as of 20 Apr 2015
  
 %\cite{Corley:2001zk}
\bibitem{Corley:2001zk} 
  S.~Corley, A.~Jevicki and S.~Ramgoolam,
  ``Exact correlators of giant gravitons from dual N=4 SYM theory,''
  Adv.\ Theor.\ Math.\ Phys.\  {\bf 5}, 809 (2002)
  [hep-th/0111222].
  %%CITATION = HEP-TH/0111222;%%
  %313 citations counted in INSPIRE as of 06 Jan 2015 



  %\cite{Caputa:2013vla}
\bibitem{Caputa:2013vla} 
  P.~Caputa, R.~d.~M.~Koch and P.~Diaz,
  ``Operators, Correlators and Free Fermions for SO(N) and Sp(N),''
  JHEP {\bf 1306}, 018 (2013)
  [arXiv:1303.7252 [hep-th]].
  %%CITATION = ARXIV:1303.7252;%%
  %7 citations counted in INSPIRE as of 06 Jan 2015
  
 %\cite{Mukhi:2005cv}
\bibitem{Mukhi:2005cv} 
  S.~Mukhi and M.~Smedback,
  ``Bubbling orientifolds,''
  JHEP {\bf 0508}, 005 (2005)
  [hep-th/0506059].
  %%CITATION = HEP-TH/0506059;%%
  %14 citations counted in INSPIRE as of 06 Jan 2015 
  
%\cite{Berenstein:2005aa}
\bibitem{Berenstein:2005aa} 
  D.~Berenstein,
 ``Large N BPS states and emergent quantum gravity,''
  JHEP {\bf 0601}, 125 (2006)
  [hep-th/0507203].
  %%CITATION = HEP-TH/0507203;%%
  %136 citations counted in INSPIRE as of 06 Jan 2015  
  

  
  
  %\cite{Berenstein:2007wi}
\bibitem{Berenstein:2007wi} 
  D.~Berenstein,
  ``Strings on conifolds from strong coupling dynamics, part I,''
  JHEP {\bf 0804}, 002 (2008)
  [arXiv:0710.2086 [hep-th]].
  %%CITATION = ARXIV:0710.2086;%%
  %27 citations counted in INSPIRE as of 06 Jan 2015
  
    %\cite{Berenstein:2007kq}
\bibitem{Berenstein:2007kq} 
  D.~E.~Berenstein and S.~A.~Hartnoll,
  ``Strings on conifolds from strong coupling dynamics: Quantitative results,''
  JHEP {\bf 0803}, 072 (2008)
  [arXiv:0711.3026 [hep-th]].
  %%CITATION = ARXIV:0711.3026;%%
  %23 citations counted in INSPIRE as of 06 Jan 2015
  
 %\cite{Berenstein:2013eya}
\bibitem{Berenstein:2013eya} 
  D.~Berenstein and E.~Dzienkowski,
  ``Open spin chains for giant gravitons and relativity,''
  JHEP {\bf 1308}, 047 (2013)
  [arXiv:1305.2394 [hep-th]].
  %%CITATION = ARXIV:1305.2394;%%
  %6 citations counted in INSPIRE as of 06 Jan 2015 
  
  
  %\cite{Berenstein:2014isa}
\bibitem{Berenstein:2014isa} 
  D.~Berenstein and E.~Dzienkowski,
  ``Giant gravitons and the emergence of geometric limits in $\beta$-deformations of ${\cal N}=4$ SYM,''
  arXiv:1408.3620 [hep-th].
  %%CITATION = ARXIV:1408.3620;%%
  %1 citations counted in INSPIRE as of 06 Jan 2015
  
  
  %\cite{Beisert:2005tm}
\bibitem{Beisert:2005tm} 
  N.~Beisert,
  ``The SU(2|2) dynamic S-matrix,''
  Adv.\ Theor.\ Math.\ Phys.\  {\bf 12}, 945 (2008)
  [hep-th/0511082].
  %%CITATION = HEP-TH/0511082;%%
  %447 citations counted in INSPIRE as of 20 Apr 2015
  
  %\cite{Dorey:2006dq}
\bibitem{Dorey:2006dq} 
  N.~Dorey,
  ``Magnon Bound States and the AdS/CFT Correspondence,''
  J.\ Phys.\ A {\bf 39}, 13119 (2006)
  [hep-th/0604175].
  %%CITATION = HEP-TH/0604175;%%
  %183 citations counted in INSPIRE as of 20 Apr 2015
    
  %\cite{Berenstein:2013md}
\bibitem{Berenstein:2013md} 
  D.~Berenstein,
 ``Giant gravitons: a collective coordinate approach,''
  Phys.\ Rev.\ D {\bf 87}, no. 12, 126009 (2013)
  [arXiv:1301.3519 [hep-th]].
  %%CITATION = ARXIV:1301.3519;%%
  %11 citations counted in INSPIRE as of 06 Jan 2015
  
 %\cite{Berenstein:2014zxa}
\bibitem{Berenstein:2014zxa} 
  D.~Berenstein,
  ``On the central charge extension of the N=4 SYM spin chain,''
  arXiv:1411.5921 [hep-th].
  %%CITATION = ARXIV:1411.5921;%% 
  
 %\cite{Filev:2014qxa}
\bibitem{Filev:2014qxa} 
  V.~G.~Filev and D.~O'Connor,
  ``Commuting Quantum Matrix Models,''
  arXiv:1408.1388 [hep-th].
  %%CITATION = ARXIV:1408.1388;%% 
  

%\cite{Klebanov:1998hh}
\bibitem{Klebanov:1998hh} 
  I.~R.~Klebanov and E.~Witten,
 ``Superconformal field theory on three-branes at a Calabi-Yau singularity,''
  Nucl.\ Phys.\ B {\bf 536}, 199 (1998)
  [hep-th/9807080].
  %%CITATION = HEP-TH/9807080;%%
  %813 citations counted in INSPIRE as of 08 Jan 2015    

  %\cite{Leigh:1995ep}
\bibitem{Leigh:1995ep} 
  R.~G.~Leigh and M.~J.~Strassler,
  ``Exactly marginal operators and duality in four-dimensional N=1 supersymmetric gauge theory,''
  Nucl.\ Phys.\ B {\bf 447}, 95 (1995)
  [hep-th/9503121].
  %%CITATION = HEP-TH/9503121;%%
  %490 citations counted in INSPIRE as of 26 Jan 2015



%\cite{Douglas:1998xa}
\bibitem{Douglas:1998xa} 
  M.~R.~Douglas,
  ``D-branes and discrete torsion,''
  hep-th/9807235.
  %%CITATION = HEP-TH/9807235;%%
  %133 citations counted in INSPIRE as of 27 gen 2015
  
    %\cite{Douglas:1999hq}
\bibitem{Douglas:1999hq} 
  M.~R.~Douglas and B.~Fiol,
  ``D-branes and discrete torsion. 2.,''
  JHEP {\bf 0509}, 053 (2005)
  [hep-th/9903031].
  %%CITATION = HEP-TH/9903031;%%
  %135 citations counted in INSPIRE as of 27 gen 2015
  
  %\cite{Berenstein:2000hy}
\bibitem{Berenstein:2000hy} 
  D.~Berenstein and R.~G.~Leigh,
  ``Discrete torsion, AdS / CFT and duality,''
  JHEP {\bf 0001}, 038 (2000)
  [hep-th/0001055].
  %%CITATION = HEP-TH/0001055;%%
  %90 citations counted in INSPIRE as of 07 Jan 2015  
  
  %\cite{Benvenuti:2005wi}
\bibitem{Benvenuti:2005wi} 
  S.~Benvenuti and A.~Hanany,
  ``Conformal manifolds for the conifold and other toric field theories,''
  JHEP {\bf 0508}, 024 (2005)
  [hep-th/0502043].
  %%CITATION = HEP-TH/0502043;%%
  %47 citations counted in INSPIRE as of 19 Jan 2015
  
  %\cite{Dasgupta:2000hn}
\bibitem{Dasgupta:2000hn} 
  K.~Dasgupta, S.~Hyun, K.~Oh and R.~Tatar,
 ``Conifolds with discrete torsion and noncommutativity,''
  JHEP {\bf 0009}, 043 (2000)
  [hep-th/0008091].
  %%CITATION = HEP-TH/0008091;%%
  %23 citations counted in INSPIRE as of 04 Feb 2015
  
  
  %\cite{Butti:2007aq}
\bibitem{Butti:2007aq} 
  A.~Butti, D.~Forcella, L.~Martucci, R.~Minasian, M.~Petrini and A.~Zaffaroni,
  ``On the geometry and the moduli space of beta-deformed quiver gauge theories,''
  JHEP {\bf 0807}, 053 (2008)
  [arXiv:0712.1215 [hep-th]].
  %%CITATION = ARXIV:0712.1215;%%
  %26 citations counted in INSPIRE as of 04 Feb 2015
  
  %\cite{Aharony:2015zea}
\bibitem{Aharony:2015zea} 
  O.~Aharony, M.~Berkooz and S.~J.~Rey,
  ``Rigid Holography and Six-Dimensional N=(2,0) Theories on $AdS_5 \times S^1$,''
  arXiv:1501.02904 [hep-th].
  %%CITATION = ARXIV:1501.02904;%%
  
 %\cite{Gadde:2010ku}
\bibitem{Gadde:2010ku} 
  A.~Gadde and L.~Rastelli,
  ``Twisted Magnons,''
  JHEP {\bf 1204}, 053 (2012)
  [arXiv:1012.2097 [hep-th]].
  %%CITATION = ARXIV:1012.2097;%%
  %12 citations counted in INSPIRE as of 19 Jan 2015 
 
 
  %\cite{Cachazo:2002ry}
\bibitem{Cachazo:2002ry} 
  F.~Cachazo, M.~R.~Douglas, N.~Seiberg and E.~Witten,
  ``Chiral rings and anomalies in supersymmetric gauge theory,''
  JHEP {\bf 0212}, 071 (2002)
  [hep-th/0211170].
  %%CITATION = HEP-TH/0211170;%%
  %328 citations counted in INSPIRE as of 19 Jan 2015  
  
 %\cite{Douglas:1996sw}
\bibitem{Douglas:1996sw} 
  M.~R.~Douglas and G.~W.~Moore,
  ``D-branes, quivers, and ALE instantons,''
  hep-th/9603167.
  %%CITATION = HEP-TH/9603167;%%
  %981 citations counted in INSPIRE as of 06 Jan 2015 
  
 %\cite{Kachru:1998ys}
\bibitem{Kachru:1998ys} 
  S.~Kachru and E.~Silverstein,
  ``4-D conformal theories and strings on orbifolds,''
  Phys.\ Rev.\ Lett.\  {\bf 80}, 4855 (1998)
  [hep-th/9802183].
  %%CITATION = HEP-TH/9802183;%%
  %586 citations counted in INSPIRE as of 06 Jan 2015 
  
  
 %\cite{Hanany:1998sd}
\bibitem{Hanany:1998sd} 
  A.~Hanany and Y.~H.~He,
  ``NonAbelian finite gauge theories,''
  JHEP {\bf 9902}, 013 (1999)
  [hep-th/9811183].
  %%CITATION = HEP-TH/9811183;%%
  %81 citations counted in INSPIRE as of 19 Jan 2015 
  

  
  
  
  %\cite{Berenstein:2001jr}
\bibitem{Berenstein:2001jr} 
  D.~Berenstein and R.~G.~Leigh,
  ``Resolution of stringy singularities by noncommutative algebras,''
  JHEP {\bf 0106}, 030 (2001)
  [hep-th/0105229].
  %%CITATION = HEP-TH/0105229;%%
  %50 citations counted in INSPIRE as of 06 Jan 2015
  
 %\cite{Berenstein:2009ay}
\bibitem{Berenstein:2009ay} 
  D.~Berenstein and M.~Romo,
  ``Aspects of ABJM orbifolds,''
  Adv.\ Theor.\ Math.\ Phys.\  {\bf 14} (2010)
  [arXiv:0909.2856 [hep-th]].
  %%CITATION = ARXIV:0909.2856;%%
  %11 citations counted in INSPIRE as of 06 Jan 2015 
  
%\cite{Witten:1998qj}
\bibitem{Witten:1998qj} 
  E.~Witten,
  ``Anti-de Sitter space and holography,''
  Adv.\ Theor.\ Math.\ Phys.\  {\bf 2}, 253 (1998)
  [hep-th/9802150].
  %%CITATION = HEP-TH/9802150;%%
  %6944 citations counted in INSPIRE as of 06 Jan 2015  
  
 %\cite{Gukov:1998kk}
\bibitem{Gukov:1998kk} 
  S.~Gukov,
  ``Comments on N=2 AdS orbifolds,''
  Phys.\ Lett.\ B {\bf 439}, 23 (1998)
  [hep-th/9806180].
  %%CITATION = HEP-TH/9806180;%%
  %41 citations counted in INSPIRE as of 07 Jan 2015 
  

  
  
 %\cite{Gukov:1998kn}
\bibitem{Gukov:1998kn} 
  S.~Gukov, M.~Rangamani and E.~Witten,
  ``Dibaryons, strings and branes in AdS orbifold models,''
  JHEP {\bf 9812}, 025 (1998)
  [hep-th/9811048].
  %%CITATION = HEP-TH/9811048;%%
  %72 citations counted in INSPIRE as of 07 Jan 2015 
  
 %\cite{Dey:2011ea}
\bibitem{Dey:2011ea} 
  T.~K.~Dey,
  ``Exact Large $R$-charge Correlators in ABJM Theory,''
  JHEP {\bf 1108}, 066 (2011)
  [arXiv:1105.0218 [hep-th]].
  %%CITATION = ARXIV:1105.0218;%%
  %12 citations counted in INSPIRE as of 07 Jan 2015 
  
  
 %\cite{deMelloKoch:2012kv}
\bibitem{deMelloKoch:2012kv} 
  R.~de Mello Koch, B.~A.~E.~Mohammed, J.~Murugan and A.~Prinsloo,
  ``Beyond the Planar Limit in ABJM,''
  JHEP {\bf 1205}, 037 (2012)
  [arXiv:1202.4925 [hep-th]].
  %%CITATION = ARXIV:1202.4925;%%
  %9 citations counted in INSPIRE as of 27 Jan 2015 
  
  
 %\cite{Caputa:2012dg}
\bibitem{Caputa:2012dg} 
  P.~Caputa and B.~A.~E.~Mohammed,
 ``From Schurs to Giants in ABJ(M),''
  JHEP {\bf 1301}, 055 (2013)
  [arXiv:1210.7705 [hep-th]].
  %%CITATION = ARXIV:1210.7705;%%
  %15 citations counted in INSPIRE as of 06 Jan 2015 
 
 %\cite{Pasukonis:2013ts}
\bibitem{Pasukonis:2013ts} 
  J.~Pasukonis and S.~Ramgoolam,
  ``Quivers as Calculators: Counting, Correlators and Riemann Surfaces,''
  JHEP {\bf 1304}, 094 (2013)
  [arXiv:1301.1980 [hep-th]].
  %%CITATION = ARXIV:1301.1980;%%
  %14 citations counted in INSPIRE as of 27 Jan 2015
 
 %\cite{Berenstein:2004hw}
\bibitem{Berenstein:2004hw} 
  D.~Berenstein,
  ``A Matrix model for a quantum Hall droplet with manifest particle-hole symmetry,''
  Phys.\ Rev.\ D {\bf 71}, 085001 (2005)
  [hep-th/0409115].
  %%CITATION = HEP-TH/0409115;%%
  %22 citations counted in INSPIRE as of 07 Jan 2015
  
  
  %\cite{Hanany:1997tb}
\bibitem{Hanany:1997tb} 
  A.~Hanany and A.~Zaffaroni,
  ``On the realization of chiral four-dimensional gauge theories using branes,''
  JHEP {\bf 9805}, 001 (1998)
  [hep-th/9801134].
  %%CITATION = HEP-TH/9801134;%%
  %109 citations counted in INSPIRE as of 08 Jan 2015
  
  %\cite{Hanany:1998it}
\bibitem{Hanany:1998it} 
  A.~Hanany and A.~M.~Uranga,
  ``Brane boxes and branes on singularities,''
  JHEP {\bf 9805}, 013 (1998)
  [hep-th/9805139].
  %%CITATION = HEP-TH/9805139;%%
  %111 citations counted in INSPIRE as of 08 Jan 2015
  
 
 \bibitem{QR}
  C.~P.Boyer, K. Galicki,
  ``New Einstein Metrics in Dimension Five", J. Differential Geom. 
  {\bf 57} (2001), no. 3, 485-524
  [math.DG/0003174]
  
  
  %\cite{Gauntlett:2004yd}
\bibitem{Gauntlett:2004yd} 
  J.~P.~Gauntlett, D.~Martelli, J.~Sparks and D.~Waldram,
  ``Sasaki-Einstein metrics on S**2 x S**3,''
  Adv.\ Theor.\ Math.\ Phys.\  {\bf 8}, 711 (2004)
  [hep-th/0403002].
  %%CITATION = HEP-TH/0403002;%%
  %302 citations counted in INSPIRE as of 08 Jan 2015
  
 %\cite{Martelli:2004wu}
\bibitem{Martelli:2004wu} 
  D.~Martelli and J.~Sparks,
  ``Toric geometry, Sasaki-Einstein manifolds and a new infinite class of AdS/CFT duals,''
  Commun.\ Math.\ Phys.\  {\bf 262}, 51 (2006)
  [hep-th/0411238].
  %%CITATION = HEP-TH/0411238;%%
  %194 citations counted in INSPIRE as of 08 Jan 2015 
  
 %\cite{Feng:2000mi}
\bibitem{Feng:2000mi} 
  B.~Feng, A.~Hanany and Y.~H.~He,
  ``D-brane gauge theories from toric singularities and toric duality,''
  Nucl.\ Phys.\ B {\bf 595}, 165 (2001)
  [hep-th/0003085].
  %%CITATION = HEP-TH/0003085;%%
  %185 citations counted in INSPIRE as of 08 Jan 2015 
 
%\cite{Benvenuti:2004dy}
\bibitem{Benvenuti:2004dy} 
  S.~Benvenuti, S.~Franco, A.~Hanany, D.~Martelli and J.~Sparks,
  ``An Infinite family of superconformal quiver gauge theories with Sasaki-Einstein duals,''
  JHEP {\bf 0506}, 064 (2005)
  [hep-th/0411264].
  %%CITATION = HEP-TH/0411264;%%
  %197 citations counted in INSPIRE as of 08 Jan 2015 

%\cite{Franco:2005rj}
\bibitem{Franco:2005rj} 
  S.~Franco, A.~Hanany, K.~D.~Kennaway, D.~Vegh and B.~Wecht,
 ``Brane dimers and quiver gauge theories,''
  JHEP {\bf 0601}, 096 (2006)
  [hep-th/0504110].
  %%CITATION = HEP-TH/0504110;%%
  %213 citations counted in INSPIRE as of 08 Jan 2015

%\cite{Franco:2005sm}
\bibitem{Franco:2005sm} 
  S.~Franco, A.~Hanany, D.~Martelli, J.~Sparks, D.~Vegh and B.~Wecht,
  ``Gauge theories from toric geometry and brane tilings,''
  JHEP {\bf 0601}, 128 (2006)
  [hep-th/0505211].
  %%CITATION = HEP-TH/0505211;%%
  %206 citations counted in INSPIRE as of 08 Jan 2015 
  
  %\cite{Benvenuti:2005cz}
\bibitem{Benvenuti:2005cz} 
  S.~Benvenuti and M.~Kruczenski,
  ``Semiclassical strings in Sasaki-Einstein manifolds and long operators in N=1 gauge theories,''
  JHEP {\bf 0610}, 051 (2006)
  [hep-th/0505046].
  %%CITATION = HEP-TH/0505046;%%
  %62 citations counted in INSPIRE as of 17 Apr 2015
  
  
  %\cite{Argyres:1999xu}
\bibitem{Argyres:1999xu} 
  P.~C.~Argyres, K.~A.~Intriligator, R.~G.~Leigh and M.~J.~Strassler,
  ``On inherited duality in N=1 d = 4 supersymmetric gauge theories,''
  JHEP {\bf 0004}, 029 (2000)
  [hep-th/9910250].
  %%CITATION = HEP-TH/9910250;%%
  %16 citations counted in INSPIRE as of 14 Apr 2015
  
  %\cite{Seiberg:1994pq}
\bibitem{Seiberg:1994pq} 
  N.~Seiberg,
  ``Electric - magnetic duality in supersymmetric nonAbelian gauge theories,''
  Nucl.\ Phys.\ B {\bf 435}, 129 (1995)
  [hep-th/9411149].
  %%CITATION = HEP-TH/9411149;%%
  %1212 citations counted in INSPIRE as of 16 Feb 2015

%\cite{Green:2010da}
\bibitem{Green:2010da} 
  D.~Green, Z.~Komargodski, N.~Seiberg, Y.~Tachikawa and B.~Wecht,
  ``Exactly Marginal Deformations and Global Symmetries,''
  JHEP {\bf 1006}, 106 (2010)
  [arXiv:1005.3546 [hep-th]].
  %%CITATION = ARXIV:1005.3546;%%
  %47 citations counted in INSPIRE as of 17 Feb 2015
  
  %\cite{Imeroni:2006rb}
\bibitem{Imeroni:2006rb} 
  E.~Imeroni and A.~Naqvi,
  ``Giants and loops in beta-deformed theories,''
  JHEP {\bf 0703}, 034 (2007)
  [hep-th/0612032].
  %%CITATION = HEP-TH/0612032;%%
  %18 citations counted in INSPIRE as of 20 Apr 2015
  
  
  %\cite{Witten:1979kh}
\bibitem{Witten:1979kh} 
  E.~Witten,
  ``Baryons in the 1/n Expansion,''
  Nucl.\ Phys.\ B {\bf 160}, 57 (1979).
  %%CITATION = NUPHA,B160,57;%%
  %2056 citations counted in INSPIRE as of 09 Jan 2015
  
  %\cite{Berenstein:2003ah}
\bibitem{Berenstein:2003ah} 
  D.~Berenstein,
  ``Shape and holography: Studies of dual operators to giant gravitons,''
  Nucl.\ Phys.\ B {\bf 675}, 179 (2003)
  [hep-th/0306090].
  %%CITATION = HEP-TH/0306090;%%
  %59 citations counted in INSPIRE as of 09 Jan 2015
  
 %\cite{Lee:1998bxa}
\bibitem{Lee:1998bxa} 
  S.~Lee, S.~Minwalla, M.~Rangamani and N.~Seiberg,
 ``Three point functions of chiral operators in D = 4, N=4 SYM at large N,''
  Adv.\ Theor.\ Math.\ Phys.\  {\bf 2}, 697 (1998)
  [hep-th/9806074].
  %%CITATION = HEP-TH/9806074;%%
  %358 citations counted in INSPIRE as of 09 Jan 2015 
  
 %\cite{D'Hoker:1999ea}
\bibitem{D'Hoker:1999ea} 
  E.~D'Hoker, D.~Z.~Freedman, S.~D.~Mathur, A.~Matusis and L.~Rastelli,
  ``Extremal correlators in the AdS / CFT correspondence,''
  In *Shifman, M.A. (ed.): The many faces of the superworld* 332-360
  [hep-th/9908160].
  %%CITATION = HEP-TH/9908160;%%
  %121 citations counted in INSPIRE as of 09 Jan 2015 
  
 %\cite{Skenderis:2006uy}
\bibitem{Skenderis:2006uy} 
  K.~Skenderis and M.~Taylor,
  ``Kaluza-Klein holography,''
  JHEP {\bf 0605}, 057 (2006)
  [hep-th/0603016].
  %%CITATION = HEP-TH/0603016;%%
  %54 citations counted in INSPIRE as of 02 Feb 2015 
  
%\cite{Caldarelli:2004mz}
\bibitem{Caldarelli:2004mz} 
  M.~M.~Caldarelli, D.~Klemm and P.~J.~Silva,
  ``Chronology protection in anti-de Sitter,''
  Class.\ Quant.\ Grav.\  {\bf 22}, 3461 (2005)
  [hep-th/0411203].
  %%CITATION = HEP-TH/0411203;%%
  %49 citations counted in INSPIRE as of 17 Apr 2015
  
 %\cite{Intriligator:2003jj}
\bibitem{Intriligator:2003jj} 
  K.~A.~Intriligator and B.~Wecht,
  ``The Exact superconformal R symmetry maximizes a,''
  Nucl.\ Phys.\ B {\bf 667}, 183 (2003)
  [hep-th/0304128].
  %%CITATION = HEP-TH/0304128;%%
  %299 citations counted in INSPIRE as of 09 Jan 2015 
  
  %\cite{Beasley:2001zp}
\bibitem{Beasley:2001zp} 
  C.~E.~Beasley and M.~R.~Plesser,
  ``Toric duality is Seiberg duality,''
  JHEP {\bf 0112}, 001 (2001)
  [hep-th/0109053].
  %%CITATION = HEP-TH/0109053;%%
  %121 citations counted in INSPIRE as of 09 Jan 2015
  
 %\cite{Feng:2001bn}
\bibitem{Feng:2001bn} 
  B.~Feng, A.~Hanany, Y.~H.~He and A.~M.~Uranga,
  ``Toric duality as Seiberg duality and brane diamonds,''
  JHEP {\bf 0112}, 035 (2001)
  [hep-th/0109063].
  %%CITATION = HEP-TH/0109063;%%
  %120 citations counted in INSPIRE as of 09 Jan 2015 
  
 %\cite{Itzykson:1979fi}
\bibitem{Itzykson:1979fi} 
  C.~Itzykson and J.~B.~Zuber,
  ``The Planar Approximation. 2.,''
  J.\ Math.\ Phys.\  {\bf 21}, 411 (1980).
  %%CITATION = JMAPA,21,411;%%
  %451 citations counted in INSPIRE as of 20 Apr 2015 
  
 %\cite{Gross:1989aw}
\bibitem{Gross:1989aw} 
  D.~J.~Gross and A.~A.~Migdal,
  ``A Nonperturbative Treatment of Two-dimensional Quantum Gravity,''
  Nucl.\ Phys.\ B {\bf 340}, 333 (1990).
  %%CITATION = NUPHA,B340,333;%%
  %404 citations counted in INSPIRE as of 20 Apr 2015 
  
 %\cite{Klebanov:1991qa}
\bibitem{Klebanov:1991qa} 
  I.~R.~Klebanov,
  ``String theory in two-dimensions,''
  In *Trieste 1991, Proceedings, String theory and quantum gravity '91* 30-101 and Princeton Univ. - PUPT-1271 (91/07,rec.Oct.) 72 p
  [hep-th/9108019].
  %%CITATION = HEP-TH/9108019;%%
  %226 citations counted in INSPIRE as of 27 Jan 2015 
  
  %\cite{Chau:1997pr}
\bibitem{Chau:1997pr} 
  L.~L.~Chau and O.~Zaboronsky,
  ``On the structure of the correlation functions in the normal matrix model,''
  Commun.\ Math.\ Phys.\  {\bf 196}, 203 (1998)
  [hep-th/9711091].
  %%CITATION = HEP-TH/9711091;%%
  %26 citations counted in INSPIRE as of 27 gen 2015
  
  %\cite{Alexandrov:2003qk}
\bibitem{Alexandrov:2003qk} 
  S.~Y.~Alexandrov, V.~A.~Kazakov and I.~K.~Kostov,
  ``2-D string theory as normal matrix model,''
  Nucl.\ Phys.\ B {\bf 667}, 90 (2003)
  [hep-th/0302106].
  %%CITATION = HEP-TH/0302106;%%
  %45 citations counted in INSPIRE as of 27 gen 2015
 
 %\cite{Teodorescu:2004qm}
\bibitem{Teodorescu:2004qm} 
  R.~Teodorescu, E.~Bettelheim, O.~Agam, A.~Zabrodin and P.~Wiegmann,
  ``Normal random matrix ensemble as a growth problem: Evolution of the spectral curve,''
  Nucl.\ Phys.\ B {\bf 704}, 407 (2005)
  [hep-th/0401165].
  %%CITATION = HEP-TH/0401165;%%
  %23 citations counted in INSPIRE as of 27 Jan 2015 
  
 %\cite{McGreevy:2000cw}
\bibitem{McGreevy:2000cw} 
  J.~McGreevy, L.~Susskind and N.~Toumbas,
  ``Invasion of the giant gravitons from Anti-de Sitter space,''
  JHEP {\bf 0006}, 008 (2000)
  [hep-th/0003075].
  %%CITATION = HEP-TH/0003075;%%
  %431 citations counted in INSPIRE as of 13 Jan 2015 
  
  
  %\cite{Grisaru:2000zn}
\bibitem{Grisaru:2000zn} 
  M.~T.~Grisaru, R.~C.~Myers and O.~Tafjord,
  ``SUSY and goliath,''
  JHEP {\bf 0008}, 040 (2000)
  [hep-th/0008015].
  %%CITATION = HEP-TH/0008015;%%
  %270 citations counted in INSPIRE as of 13 Jan 2015  
  
  %\cite{Hashimoto:2000zp}
\bibitem{Hashimoto:2000zp} 
  A.~Hashimoto, S.~Hirano and N.~Itzhaki,
  ``Large branes in AdS and their field theory dual,''
  JHEP {\bf 0008}, 051 (2000)
  [hep-th/0008016].
  %%CITATION = HEP-TH/0008016;%%
  %263 citations counted in INSPIRE as of 13 Jan 2015
  
  
  %\cite{Mikhailov:2000ya}
\bibitem{Mikhailov:2000ya} 
  A.~Mikhailov,
  ``Giant gravitons from holomorphic surfaces,''
  JHEP {\bf 0011}, 027 (2000)
  [hep-th/0010206].
  %%CITATION = HEP-TH/0010206;%%
  %112 citations counted in INSPIRE as of 17 Apr 2015
  
  
  %\cite{Martelli:2006vh}
\bibitem{Martelli:2006vh} 
  D.~Martelli and J.~Sparks,
 ``Dual Giant Gravitons in Sasaki-Einstein Backgrounds,''
  Nucl.\ Phys.\ B {\bf 759}, 292 (2006)
  [hep-th/0608060].
  %%CITATION = HEP-TH/0608060;%%
  %50 citations counted in INSPIRE as of 17 Apr 2015
  
  %\cite{de Mello Koch:2007uv}
\bibitem{de Mello Koch:2007uv} 
  R.~de Mello Koch, J.~Smolic and M.~Smolic,
  ``Giant Gravitons - with Strings Attached (II),''
  JHEP {\bf 0709}, 049 (2007)
  [hep-th/0701067].
  %%CITATION = HEP-TH/0701067;%%
  %67 citations counted in INSPIRE as of 17 Apr 2015
  
  %\cite{Caldarelli:2004ig}
\bibitem{Caldarelli:2004ig} 
  M.~M.~Caldarelli and P.~J.~Silva,
  ``Giant gravitons in AdS/CFT (I): Matrix model and back reaction,''
  JHEP {\bf 0408}, 029 (2004)
  [hep-th/0406096].
  %%CITATION = HEP-TH/0406096;%%
  %43 citations counted in INSPIRE as of 28 gen 2015
  
  
  
  
  %\cite{Skenderis:2007yb}
\bibitem{Skenderis:2007yb} 
  K.~Skenderis and M.~Taylor,
  ``Anatomy of bubbling solutions,''
  JHEP {\bf 0709}, 019 (2007)
  [arXiv:0706.0216 [hep-th]].
  %%CITATION = ARXIV:0706.0216;%%
  %29 citations counted in INSPIRE as of 13 Jan 2015




%\cite{Berenstein:2002ge}
\bibitem{Berenstein:2002ge} 
  D.~Berenstein,
  ``Reverse geometric engineering of singularities,''
  JHEP {\bf 0204}, 052 (2002)
  [hep-th/0201093].
  %%CITATION = HEP-TH/0201093;%%
  %40 citations counted in INSPIRE as of 21 Jan 2015



%\cite{Diaconescu:1997br}
\bibitem{Diaconescu:1997br} 
  D.~E.~Diaconescu, M.~R.~Douglas and J.~Gomis,
 ``Fractional branes and wrapped branes,''
  JHEP {\bf 9802}, 013 (1998)
  [hep-th/9712230].
  %%CITATION = HEP-TH/9712230;%%
  %196 citations counted in INSPIRE as of 20 Apr 2015


%\cite{Blau:2002dy}
\bibitem{Blau:2002dy} 
  M.~Blau, J.~M.~Figueroa-O'Farrill, C.~Hull and G.~Papadopoulos,
  ``Penrose limits and maximal supersymmetry,''
  Class.\ Quant.\ Grav.\  {\bf 19}, L87 (2002)
  [hep-th/0201081].
  %%CITATION = HEP-TH/0201081;%%
  %448 citations counted in INSPIRE as of 13 Jan 2015
  
  
 %\cite{Berenstein:2002jq}
\bibitem{Berenstein:2002jq} 
  D.~E.~Berenstein, J.~M.~Maldacena and H.~S.~Nastase,
  ``Strings in flat space and pp waves from N=4 superYang-Mills,''
  JHEP {\bf 0204}, 013 (2002)
  [hep-th/0202021].
  %%CITATION = HEP-TH/0202021;%%
  %1469 citations counted in INSPIRE as of 13 Jan 2015 



%\cite{Itzhaki:2002kh}
\bibitem{Itzhaki:2002kh} 
  N.~Itzhaki, I.~R.~Klebanov and S.~Mukhi,
  ``PP wave limit and enhanced supersymmetry in gauge theories,''
  JHEP {\bf 0203}, 048 (2002)
  [hep-th/0202153].
  %%CITATION = HEP-TH/0202153;%%
  %116 citations counted in INSPIRE as of 13 Jan 2015

%\cite{Gomis:2002km}
\bibitem{Gomis:2002km} 
  J.~Gomis and H.~Ooguri,
  ``Penrose limit of N = 1 gauge theories,''
  Nucl.\ Phys.\ B {\bf 635}, 106 (2002)
  [hep-th/0202157].
  %%CITATION = HEP-TH/0202157;%%
  %134 citations counted in INSPIRE as of 13 Jan 2015
  
 %\cite{Pando Zayas:2002rx}
\bibitem{Pando Zayas:2002rx} 
  L.~A.~Pando Zayas and J.~Sonnenschein,
  ``On Penrose limits and gauge theories,''
  JHEP {\bf 0205}, 010 (2002)
  [hep-th/0202186].
  %%CITATION = HEP-TH/0202186;%%
  %117 citations counted in INSPIRE as of 13 Jan 2015 
  
 %\cite{Berenstein:2002sa}
\bibitem{Berenstein:2002sa} 
  D.~Berenstein and H.~Nastase,
  ``On light cone string field theory from superYang-Mills and holography,''
  hep-th/0205048.
  %%CITATION = HEP-TH/0205048;%%
  %153 citations counted in INSPIRE as of 13 Jan 2015 
  
  
 %\cite{Berenstein:2000ux}
\bibitem{Berenstein:2000ux} 
  D.~Berenstein, V.~Jejjala and R.~G.~Leigh,
  ``Marginal and relevant deformations of N=4 field theories and noncommutative moduli spaces of vacua,''
  Nucl.\ Phys.\ B {\bf 589}, 196 (2000)
  [hep-th/0005087].
  %%CITATION = HEP-TH/0005087;%%
  %107 citations counted in INSPIRE as of 26 Jan 2015 
  
  

  
 %\cite{Brezin:1977sv}
\bibitem{Brezin:1977sv} 
  E.~Brezin, C.~Itzykson, G.~Parisi and J.~B.~Zuber,
  ``Planar Diagrams,''
  Commun.\ Math.\ Phys.\  {\bf 59}, 35 (1978).
  %%CITATION = CMPHA,59,35;%%
  %1053 citations counted in INSPIRE as of 28 gen 2015 
  
  
  %\cite{Seiberg:1994pq}
\bibitem{Seiberg:1994pq} 
  N.~Seiberg,
  ``Electric - magnetic duality in supersymmetric nonAbelian gauge theories,''
  Nucl.\ Phys.\ B {\bf 435}, 129 (1995)
  [hep-th/9411149].
  %%CITATION = HEP-TH/9411149;%%
  %1229 citations counted in INSPIRE as of 17 Apr 2015
  
  %\cite{Howe:1998zi}
\bibitem{Howe:1998zi} 
  P.~S.~Howe, E.~Sokatchev and P.~C.~West,
  ``Three point functions in N=4 Yang-Mills,''
  Phys.\ Lett.\ B {\bf 444}, 341 (1998)
  [hep-th/9808162].
  %%CITATION = HEP-TH/9808162;%%
  %88 citations counted in INSPIRE as of 26 Jan 2015
  
  %\cite{Heslop:2002hp}
\bibitem{Heslop:2002hp} 
  P.~J.~Heslop and P.~S.~Howe,
  ``Four point functions in N=4 SYM,''
  JHEP {\bf 0301}, 043 (2003)
  [hep-th/0211252].
  %%CITATION = HEP-TH/0211252;%%
  %31 citations counted in INSPIRE as of 26 Jan 2015
  
  %\cite{Dolan:2004mu}
\bibitem{Dolan:2004mu} 
  F.~A.~Dolan, L.~Gallot and E.~Sokatchev,
  ``On four-point functions of 1/2-BPS operators in general dimensions,''
  JHEP {\bf 0409}, 056 (2004)
  [hep-th/0405180].
  %%CITATION = HEP-TH/0405180;%%
  %24 citations counted in INSPIRE as of 26 Jan 2015
  
  %\cite{David:2013oha}
\bibitem{David:2013oha} 
  J.~R.~David and A.~Sadhukhan,
  ``Structure constants of $\beta$ deformed super Yang-Mills,''
  JHEP {\bf 2013}, 206 (2013)
  [arXiv:1307.3909 [hep-th]].
  %%CITATION = ARXIV:1307.3909;%%
  %1 citations counted in INSPIRE as of 26 Jan 2015
  

  
 %\cite{Young:2014lka}
\bibitem{Young:2014lka} 
  D.~Young,
  ``ABJ(M) Chiral Primary Three-Point Function at Two-loops,''
  JHEP {\bf 1407}, 120 (2014)
  [arXiv:1404.1117 [hep-th]].
  %%CITATION = ARXIV:1404.1117;%%
  %1 citations counted in INSPIRE as of 27 gen 2015 
  
   %\cite{Aharony:2008ug}
\bibitem{Aharony:2008ug} 
  O.~Aharony, O.~Bergman, D.~L.~Jafferis and J.~Maldacena,
  ``N=6 superconformal Chern-Simons-matter theories, M2-branes and their gravity duals,''
  JHEP {\bf 0810}, 091 (2008)
  [arXiv:0806.1218 [hep-th]].
  %%CITATION = ARXIV:0806.1218;%%
  %1221 citations counted in INSPIRE as of 27 gen 2015 
  
 %\cite{Dijkgraaf:2002dh}
\bibitem{Dijkgraaf:2002dh} 
  R.~Dijkgraaf and C.~Vafa,
  ``A Perturbative window into nonperturbative physics,''
  hep-th/0208048.
  %%CITATION = HEP-TH/0208048;%%
  %485 citations counted in INSPIRE as of 26 Jan 2015 
  
  
 %\cite{Berenstein:2003fx}
\bibitem{Berenstein:2003fx} 
  D.~Berenstein,
  ``D-brane realizations of runaway behavior and moduli stabilization,''
  hep-th/0303230.
  %%CITATION = HEP-TH/0303230;%%
  %11 citations counted in INSPIRE as of 27 gen 2015 
  
%\cite{Dorey:2002pq}
\bibitem{Dorey:2002pq} 
  N.~Dorey, T.~J.~Hollowood and S.~P.~Kumar,
  ``S duality of the Leigh-Strassler deformation via matrix models,''
  JHEP {\bf 0212}, 003 (2002)
  [hep-th/0210239].
  %%CITATION = HEP-TH/0210239;%%
  %82 citations counted in INSPIRE as of 20 Apr 2015
  
  
%\cite{Arik:1973vg}
\bibitem{Arik:1973vg} 
  M.~Arik and D.~D.~Coon,
``Hilbert Spaces of Analytic Functions and Generalized Coherent States,''
  J.\ Math.\ Phys.\  {\bf 17}, 524 (1976).
  %%CITATION = JMAPA,17,524;%%
  %179 citations counted in INSPIRE as of 12 Jan 2015  
  
 %\cite{'tHooft:1973jz}
\bibitem{'tHooft:1973jz} 
  G.~'t Hooft,
``A Planar Diagram Theory for Strong Interactions,''
  Nucl.\ Phys.\ B {\bf 72}, 461 (1974).
  %%CITATION = NUPHA,B72,461;%%
  %3721 citations counted in INSPIRE as of 27 Jan 2015 
  
\end{thebibliography}
\end{document}